\documentclass[twocolumn,twocolappendix]{aastex631}

\usepackage{changepage} 
\usepackage{rotating}

\usepackage{tabularx} 
\usepackage{scrextend}
\let\orgautoref\autoref
\renewcommand{\autoref}
        {\def\equationautorefname{Eq.}%
         \def\figureautorefname{Fig.}%
         \def\sectionautorefname{Sect.}%
         \def\subsectionautorefname{Sect.}%
         \def\subsubsectionautorefname{Sect.}%
         \orgautoref}

\usepackage{xcolor}
\definecolor{dark-red}{rgb}{0.9,0.0,0.0}
\definecolor{dark-blue}{rgb}{0.15,0.15,0.9}
\definecolor{dark-green}{rgb}{0.15,0.8,0.15}
\definecolor{medium-blue}{rgb}{0,0,0.9}
\hypersetup{linkcolor={dark-blue},citecolor={dark-blue}, urlcolor={medium-blue}
}

\usepackage[flushleft]{threeparttable}

\begin{document}

\title{TOI-4504: Exceptionally
large Transit Timing Variations \\ induced by two resonant warm gas giants in a three planet system.}

\correspondingauthor{Michaela V\'{i}tkov\'{a}}
\email{vitkova@asu.cas.cz}

\author[0000-0002-2994-2929]{Michaela V\'{i}tkov\'{a}}
\affiliation{Astronomical Institute of the Czech Academy of Sciences, Fri\v{c}ova 298, CZ-25165 Ond\v{r}ejov, Czech Republic}
\affiliation{Department of Theoretical Physics and Astrophysics, Faculty of Science, Masaryk University, Kotl\'{a}\v{r}sk\'{a} 2, CZ-61137 Brno, Czech Republic}

\author[0000-0002-9158-7315]{Rafael Brahm} 
\affiliation{Facultad de Ingeniera y Ciencias, Universidad Adolfo Ib\'{a}\~{n}ez, Av. Diagonal las Torres 2640, Pe\~{n}alol\'{e}n, Santiago, Chile}
\affiliation{Millennium Institute for Astrophysics, Chile}
\affiliation{Data Observatory Foundation, Chile}

\author[0000-0002-0236-775X]{Trifon Trifonov} 
\affiliation{Max-Planck-Institut für Astronomie, Königstuhl 17, D-69117 Heidelberg, Germany}
\affiliation{Department of Astronomy, Sofia University ``St Kliment Ohridski'', 5 James Bourchier Blvd, BG-1164 Sofia, Bulgaria}
\affiliation{Landessternwarte, Zentrum f\"ur Astronomie der Universit\"at Heidelberg, K\"onigstuhl 12, D-69117 Heidelberg, Germany}

\author[0000-0002-1623-5352]{Petr Kab\'{a}th} 
\affiliation{Astronomical Institute of the Czech Academy of Sciences, Fri\v{c}ova 298, CZ-25165 Ond\v{r}ejov, Czech Republic}

\author[0000-0002-5389-3944]{Andr\'es Jord\'an} 
\affiliation{Facultad de Ingeniera y Ciencias, Universidad Adolfo Ib\'{a}\~{n}ez, Av. Diagonal las Torres 2640, Pe\~{n}alol\'{e}n, Santiago, Chile}
\affiliation{Millennium Institute for Astrophysics, Chile}
\affiliation{Data Observatory Foundation, Chile}

\author[0000-0002-1493-300X]{Thomas Henning} 
\affiliation{Max-Planck-Institut für Astronomie, Königstuhl 17, D-69117 Heidelberg, Germany}

\author[0000-0002-5945-7975]{Melissa J. Hobson} 
\affiliation{Observatoire de Genève, Département d'Astronomie, Université de Genève, Chemin Pegasi 51b, 1290 Versoix, Switzerland}

\author[0000-0003-3130-2768]{Jan Eberhardt} 
\affiliation{Max-Planck-Institut für Astronomie, Königstuhl 17, D-69117 Heidelberg, Germany}

\author[0009-0004-8891-4057]{Marcelo Tala Pinto} 
\affiliation{Facultad de Ingeniera y Ciencias, Universidad Adolfo Ib\'{a}\~{n}ez, Av. Diagonal las Torres 2640, Pe\~{n}alol\'{e}n, Santiago, Chile}

\author[0000-0003-3047-6272]{Felipe I. Rojas} 
\affiliation{Instituto de Astrof\'isica, Facultad de F\'isica, Pontificia Universidad Cat\'olica de Chile, Chile}

\author[0000-0001-9513-1449]{Nestor Espinoza} 
\affiliation{Space Telescope Science Institute, 3700 San Martin Drive, Baltimore, MD 21218, USA}

\author[0000-0001-8355-2107]{Martin Schlecker} 
\affiliation{Department of Astronomy/Steward Observatory, The University of Arizona, 933 North Cherry Avenue, Tucson, AZ 85721, USA}

\author{Mat\'{i}as I. Jones} 
\affiliation{European Southern Observatory, Alonso de Córdova 3107, Vitacura, Casilla 19001, Santiago, Chile}



\author[0000-0002-7927-9555]{Maximiliano Moyano} 
\affiliation{Instituto de Astronom\'ia, Universidad Cat\'olica del Norte, Angamos 0610, 1270709, Antofagasta, Chile}

\author[0000-0003-4723-9660]{Susana Eyheramendy} 
\affiliation{Facultad de Ingeniera y Ciencias, Universidad Adolfo Ib\'{a}\~{n}ez, Av. Diagonal las Torres 2640, Pe\~{n}alol\'{e}n, Santiago, Chile}
\affiliation{Millennium Institute for Astrophysics, Chile}
\affiliation{Data Observatory Foundation, Chile}

\author[0000-0002-0619-7639]{Carl Ziegler} 
\affiliation{Department of Physics, Engineering and Astronomy, Stephen F. Austin State University, 1936 North St, Nacogdoches, TX 75962, USA}


\author[0000-0001-6513-1659]{Jack J. Lissauer}  
\affiliation{NASA Ames Research Center, Moffett Field, CA 94035, USA}

\author[0000-0001-7246-5438]{Andrew Vanderburg}  
\affiliation{Department of Astronomy, University of Wisconsin-Madison, Madison, WI 53706, USA}

\author[0000-0001-6588-9574]{Karen A.\ Collins}  
\affiliation{Center for Astrophysics \textbar \ Harvard \& Smithsonian, 60 Garden Street, Cambridge, MA 02138, USA}

\author[0000-0002-5402-9613]{Bill Wohler} 
\affiliation{SETI Institute, Mountain View, CA 94043 USA/NASA Ames Research Center, Moffett Field, CA 94035 USA}

\author[0000-0002-3555-8464]{David Watanabe}
\affiliation{Planetary Discoveries, Valencia CA 91354, USA}

\author[0000-0003-2058-6662]{George R. Ricker}
\affiliation{Department of Physics and Kavli Institute for Astrophysics and Space Research, Massachusetts Institute of Technology, 77 Massachusetts Avenue, Cambridge, MA 02139, USA}

\author[0000-0001-6763-6562]{Roland Vanderspek}
\affiliation{Department of Physics and Kavli Institute for Astrophysics and Space Research, Massachusetts Institute of Technology, 77 Massachusetts Avenue, Cambridge, MA 02139, USA}


\author[0000-0002-6892-6948]{Sara Seager}
\affiliation{Department of Physics and Kavli Institute for Astrophysics and Space Research, Massachusetts Institute of Technology, 77 Massachusetts Avenue, Cambridge, MA 02139, USA}
\affiliation{Department of Earth, Atmospheric and Planetary Sciences, Massachusetts Institute of Technology, 77 Massachusetts Avenue, Cambridge, MA 02139, USA}
\affiliation{Department of Aeronautics and Astronautics, Massachusetts Institute of Technology, 77 Massachusetts Avenue, Cambridge, MA 02139, USA}

\author[0000-0002-4265-047X]{Joshua N.\ Winn}
\affiliation{Department of Astrophysical Sciences, Princeton University, Princeton, NJ 08544, USA}

\author[0000-0002-4715-9460]{Jon M. Jenkins}
\affiliation{NASA Ames Research Center, Moffett Field, CA 94035, USA}

\author[0000-0002-7602-0046]{Marek Skarka} 
\affiliation{Astronomical Institute of the Czech Academy of Sciences, Fri\v{c}ova 298, CZ-25165 Ond\v{r}ejov, Czech Republic}
\affiliation{Department of Theoretical Physics and Astrophysics, Faculty of Science, Masaryk University, Kotl\'{a}\v{r}sk\'{a} 2, CZ-61137 Brno, Czech Republic}





\begin{abstract}
We present a joint analysis of TTVs and Doppler data for the transiting exoplanet system TOI-4504. TOI-4504\,c is a warm Jupiter-mass planet that exhibits the largest known transit timing variations (TTVs), with a peak-to-node amplitude of $\sim$ 2\,days, the largest value ever observed, and a super-period of $\sim$ 930\,d. TOI-4504\,b and c were identified in public TESS data, while the TTVs observed in TOI-4504\,c, together with radial velocity (RV) data collected with FEROS, allowed us to uncover a third, non-transiting planet in this system, TOI-4504\,d. We were able to detect transits of TOI-4504\,b in the TESS data with a period of 2.4261$\pm 0.0001$\,days and derive a radius of 2.69$\pm 0.19$\,R$_{\oplus}$. The RV scatter of TOI-4504 was too large to constrain the mass of TOI-4504\,b, but the RV signals of TOI-4504\,c \& d were sufficiently large to measure their masses. The TTV+RV dynamical model we apply confirms TOI-4504\,c as a warm Jupiter planet with an osculating period of 82.54$\pm 0.02$\,d, mass of 3.77$\pm 0.18$\,M$_{\rm J}$ and a radius of 0.99$\pm 0.05$\,R$_{\rm J}$, while the non-transiting planet TOI-4504\,d, has an orbital period of 40.56$\pm 0.04$\,days and mass of 1.42$_{-0.06}^{+0.07}$\,M$_{\rm J}$. We present the discovery of a system with three exoplanets: a hot sub-Neptune and two warm Jupiter planets. The gas giant pair is stable and likely locked in a first-order 2:1 mean-motion resonance (MMR). The TOI-4504 system is an important addition to MMR pairs, whose increasing occurrence supports a smooth migration into a resonant configuration during the protoplanetary disk phase. 
\end{abstract}


\section{Introduction} \label{sec:intro}

More than 5,000 transiting exoplanets have been identified. In a Keplerian system, transits occur at regular intervals. However, if additional planets are present in a system and their gravitational interactions are significant on observable timescales, then Transit Timing Variations \citep[TTVs,][]{TTV1,TTV2} are expected over the observed baseline. In particular, planets in or near low-order mean-motion resonances (MMRs) exhibit the largest TTV signals \citep{AGOL2005}. TTVs provide important constraints on the planetary masses and orbital parameters in the system and sometimes help uncover non-transiting planets. For instance, the first non-transiting planet fully characterized through TTVs was Kepler-46\,c \citep{KEPLER46}, which induced TTVs on the transiting planet Kepler-46\,b. Another example is Kepler-88\,b with a TTV amplitude of approximately 12 hours (peak-to-node), earning it the title ``the King of TTVs" \citep{KOI-142}. These TTVs revealed a pair of planets near the 2:1 MMRs. Another case of 2:1 resonance causing the peak-to-node TTVs of $\sim$ 1 day can be seen in Kepler-30\,b \citep{Kepler30a,Kepler30b}. Kepler-90\,g has a variation of 25.7 hours in the time lapsed between consecutive transits, but only a few transits have been observed, and the full amplitude of variation is not known \citep{Kepler90}.

Currently, we know of 30 planets discovered by TTVs\footnote{NASA Exoplanet Archive, Sept 3, 2024: \url{https://exoplanetarchive.ipac.caltech.edu}}, and more are being detected by NASA’s Transiting Exoplanet Survey Satellite \citep[TESS,][]{TESS}. TESS aims to detect transiting planets around relatively bright stars that are suitable for precise radial velocity (RV) measurements and, in rare cases, could help uncover non-transiting planets through TTVs. A combination of precise RV measurements and TTVs helps determine the planetary mass, radius, bulk density, and other important physical parameters.

\begin{table}[ht!]
\caption{Stellar parameters of TOI-4504.}
\centering
\label{star_params}
\begin{tabular}{ccc}
 \hline 
  \hline
Parameter     & Value                   & Reference                        \\ \hline
Names         & TIC 349972412, TOI-4504 & TIC v8.2                             \\
RA (J2000)            & 07$^{\rm{h}}$ 37$^{\rm{m}}$ 52.1529498945$^{\rm{s}}$      & Gaia DR3                         \\
DEC (J2000)           & -62$^{\circ}$ 04$'$ 41.803583657$''$	      & Gaia DR3                         \\
T [mag]            & 12.5542                 & TIC v8.2                         \\
B [mag]            & 14.239                  & TIC v8.2                         \\
V [mag]            & 13.364                  & TIC v8.2                         \\
distance [pc]     & 342.605 ${\pm}$ 1.707      & TIC v8.2                         \\
Spectral type & K1V                     & P\&M \\
T$_{\rm{eff}}$ [K]         & 5315 ${\pm}$ 60             & this work                         \\
$\log g$ [cm s$^{-2}$]        & 4.458$^{+0.021}_{-0.015}$       & this work                         \\
R$_{\star}$ [R$_{\odot}$]            & 0.92$^{+0.04}_{-0.04}$           & this work                         \\
M$_{\star}$ [M$_{\odot}$]             & 0.89$^{+0.06}_{-0.04}$         & this work                         \\
L$_{\star}$ [L$_{\odot}$]             & 0.62$^{+0.03}_{-0.03}$       & this work                         \\
$\rho_{\star}$ [kg m$^{-3}$]          & 1607$^{+93}_{-64}$        & this work         \\ 
\,[Fe/H] [dex]         & 0.16 ${\pm}$ 0.05     & this work                         \\
$v \cdot \sin i$ [km s$^{-1}$]             & 1.9 ${\pm}$ 0.5     & this work                         \\
Age [Gyr]             & 10.0$^{+2.9}_{-3.6}$    & this work                         \\
$\rm{A_V}$ [mag]             & 0.35$^{+0.07}_{-0.06}$    & this work                         \\
\hline
\hline
\end{tabular}
\tablecomments{P\&M: \cite{sp_type} \\ TIC v8.2: \cite{tic_1,tic_2} \\ Gaia DR3: \cite{gaia_a,gaia_b}} 
 \end{table}

\begin{figure}[ht!]
  \centering
\includegraphics[width=0.48\textwidth]{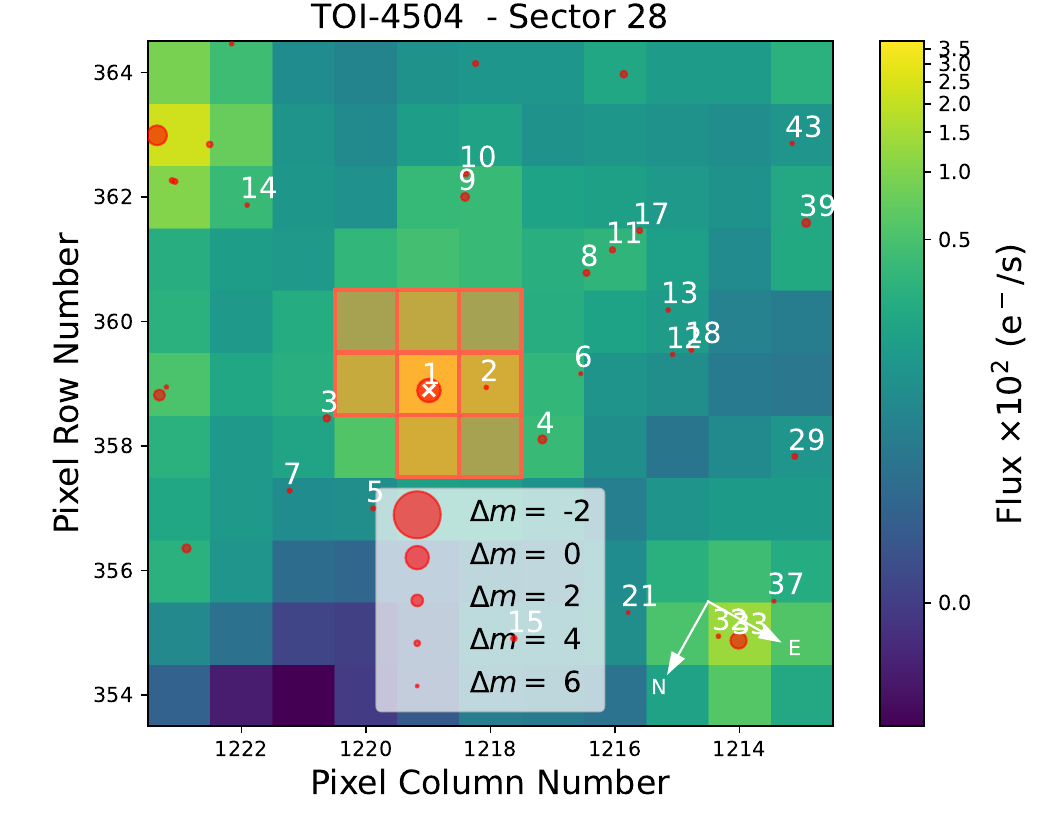}
  \caption{Target Pixel Files in Sector 28 for TOI-4504 obtained with {\textsc tpfplotter}. The shape of the aperture mask used to extract the photometry is marked with orange. Red dots indicate the sources of the Gaia DR3 catalogue in the field. TOI-4504 is marked with a white cross.}
  \label{TPF}
\end{figure}

In this work, we confirm the exoplanetary nature of the detected signal from TOI-4504. We firmly validate TOI-4504\,c, and report evidence for a non-transiting planet TOI-4504\,d that causes very strong TTVs of TOI-4504\,c. TOI-4504.01 (hereafter, TOI-4504\,c) was identified by The~Warm~gIaNts~with~tEss (WINE) collaboration, which is dedicated to the systematic characterization of TESS transiting warm giant planets \citep[see e.g.,][and references therein]{wine1,wine2,wine3,wine4,wine5,Bozhilov2023,TOI-199,jones2024}. The signature was referred to the TESS Science Office at MIT as a CTOI in May 2020, and the Quick Look Pipeline \citep[QLP,][]{QLPa,QLPb} was run to assess the candidate, identifying a period of P = 81.966 days. The TESS Science Processing Operations Center \citep[SPOC,][]{SPOC} independently detected transits of TOI-4504\,c in Sector 34 and several subsequent single- and multi-sector searches, using a noise-compensating matched filter \citep{Jenkins2002,Jenkins2020}, and an initial limb-darkened transit model was fit \citep{Li2019} and a suite of diagnostic tests were conducted to help make or break the planetary interpretation \citep{Twicken2018}. The shallower signal designated as TOI-4504.02 (hereafter TOI-4504\,b) was detected by the TESS Science Processing Operation Center \citep[SPOC,][]{SPOC} at NASA Ames Research Center with P = 2.4261 days in a search of sectors 27-36. Both TOI-4504\,b and c were alerted to the community by the TSO on 21 October 2021.


\begin{figure}[ht!]
  \centering
  \includegraphics[width=0.9\linewidth]{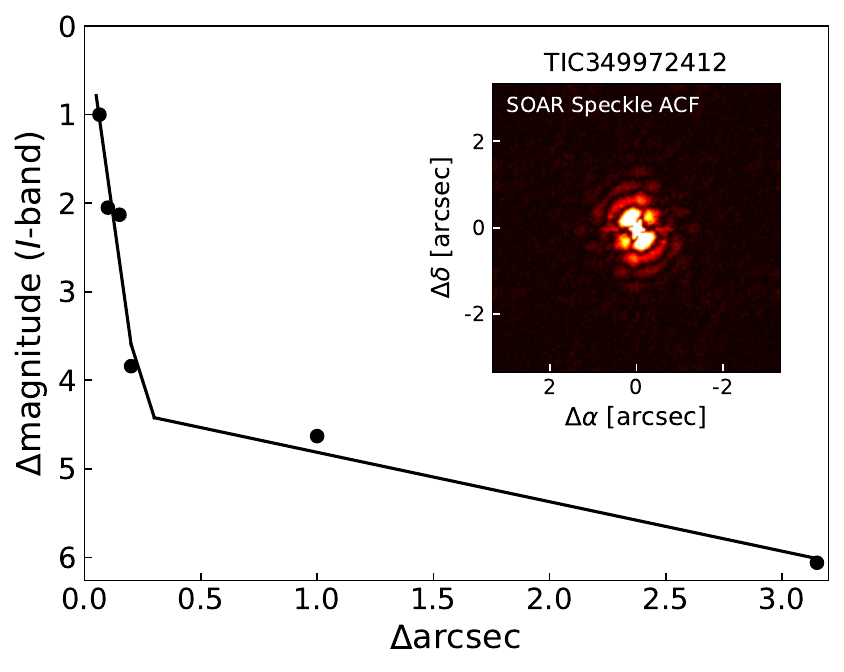}
  \caption{High-resolution imaging from SOAR for TOI-4504. The inside image shows a speckle auto-correlation function. The 5$\sigma$ contrast curve is shown as the black points with the linear fit as the black solid line.}
  \label{Imaging}
\end{figure}

In this work, we report evidence for a non-transiting planet TOI-4504\,d that causes very strong TTVs of TOI-4504\,c. The induced TTVs on TOI-4504\,c have a semi-amplitude of a little less than 2 days (peak-to-peak of 4 days), making TOI-4504\,c a new record-holder for TTV signal amplitude.
Given the strong sinusoidal TTV signal we detect, near a low-order period ratio commensurability with the transiting planet is suspected. In such cases, the TTV periodic signal can be approximated by the so-called ``super-period":

\begin{equation}
{P}_{\mathrm{TTV}}={\left|\frac{j-1}{{P}_{\mathrm{in}}}-\frac{j}{{P}_{\mathrm{out}}}\right|}^{-1}.
\end{equation}

\noindent
where $j$ = $P_{\mathrm{out}}$/$P_{\mathrm{in}}$ represents the close commensurability between the inner and the outer planets. The observed super-period of TOI-4504 is $\sim$ 930\,d. Our fitting analysis of the observed data indicates that TOI-4504\,d is an inner Jovian-mass planet with a period of about 41 days, placing it in a 2:1 period-ratio commensurability with the transiting exoplanet TOI-4504\,c.

This paper is organized as follows: In \autoref{sec:observ}, we introduce the photometric and spectroscopic data collected for TOI-4504. In \autoref{sec3}, we present the stellar parameters 
of TOI-4504. In \autoref{sec4}, we introduce our orbital analysis and results. Finally, in \autoref{sec5}, we present a summary and our conclusions.

\section{Observations} 
\label{sec:observ}

\subsection{TESS photometry}
\label{sec:photo}
TOI-4504 was observed with a 30-minute cadence during the first year of the TESS primary mission in sectors 1-13. In the third and fifth years (sectors 27-38, 61-65, and 67-69), it was observed with a 120-second cadence. The image data were reduced and analyzed by the Science Processing Operations Center at NASA Ames Research Center.

We retrieved the 30-minute data using the {\textsc tesseract} pipeline\,\footnote{Available at \url{https://github.com/astrofelipe/tesseract}}(Rojas et al. in prep.). This pipeline extracts light curves from Full Frame Images (FFIs) using {\textsc TESSCut} \citep{tesscut} and {\textsc Lightkurve} \citep{lightkurve} packages. The download of the data with a 120\,s cadence was done with {\textsc Lightkurve}. Transits of TOI-4504\,c occur in sectors 3, 6, 9, 12, 28, 31, 34, 37, 61, 64 and 67. We used the Presearch Data Conditioning SAP flux (PDCSAP, \citealt{pdcsap1,pdcsap2,pdcsap3}) in all sectors where it is available. PDCSAP light curves are corrected for systematic trends. The transit present in sector 61 is cut off by a quality check in PDCSAP. This transit can be seen in Simple Aperture Photometry (SAP) data but is contaminated by background noise. It was possible to use data from this sector for measuring TOI-4504\,c transit time, but we did not use these data to fit the planet parameters. Analysis of TOI-4504\,c is described in more detail in \autoref{sec:4504c}.

For the TOI-4504\,b analysis, we used 120\,s cadence data available in sectors 27-38, 61-65, and 67-69 and used PDCSAP light curves. Analysis of TOI-4504\,c is described in \autoref{sec:4504b}.

\subsection{Spectroscopic data}
\label{sec:spec}
Follow-up observations of TOI-4504 were performed with the FEROS spectrograph \citep{feros} mounted at the MPG 2.2m telescope at La Silla Observatory. TOI-4504 was observed between March 2020 and May 2024 in order to validate the transiting companions and constrain their masses. We obtained 39 spectra with exposure times of 1500 s and 1800 s and an average signal-to-noise ratio of 40. We used the {\textsc ceres} pipeline \citep{ceres} to process the data. From this pipeline, we obtained RV and stellar activity indicator such as CCF bisector spans (BIS, e.g., \cite{bis}), $\rm{H}_{\alpha}$ \citep{haplha}, $\log(R'_{HK})$ \citep{log,log2}, Na II, and He I \citep{he}.
Our extracted RVs and activity indices for TOI-4504 are listed in \autoref{RV_feros}.

\subsection{Inspection of nearby sources}
\label{sec:contaminant}
For inspection of Target Pixel Files (TPF), we used {\textsc tpfplotter} \citep{tpfplotter}. It compares the {\it Gaia} DR3 catalogue with TESS TPF and allows us to see possible contamination in the field. In \autoref{TPF}, a TPF image of sector 28 is shown. In almost all sectors, a star with {\it Gaia} ID 5289275864525442048 (TIC 349972416, $G=18.48$\,mag, star '2' in \autoref{TPF}) is in the aperture mask. It is 5.4 magnitudes fainter than our object, and the distance between our target and TIC 349972416 is 19.37\,$''$. Among the multiple sector data, there are two additional sources, '3' and '4', which are often close or in the aperture; however, if the transit events were on the possible contaminator stars, the transit depth would vary substantially in the observed sectors, which were not observed. Our RVs show that the transits likely originate from the target star and not from this companion (\autoref{sec:spec}).

To reject the contamination by sources closer than $\sim$2 \,$''$ from the target star, we used the 4.1-m SOAR telescope \citep{SOAR} within the SOAR TESS survey \citep{SOAR_TESS} to obtain speckle imaging. The images were obtained on 15 April 2022 with a Cousins I filter and a resolution of 36 mas, and they did not reveal any nearby sources. In \autoref{Imaging}, the speckle auto-correlation function and the contrast curve are shown. It reaches a contrast of $\Delta$mag = 4.5 at 0.5\,arcsec, and an estimated point spread function (PSF) is 0.064\,arcsec. There is no apparent nearby contaminant within 3\,arcsec from the target (\autoref{Imaging}).
Furthermore, the Gaia Renormalised Unit Weight Error (RUWE) value 1.12 of our object also indicates that the single-star model fits the astrometric observations well.
Additionally, the SPOC difference image centroid test was able to localize the host star to within 0.33 $\pm$ 2.7\,arcsec of the transit source location (averaged over the different TCEs in the S27-69 search) for TOI-4504\,c, and to 1.8 $\pm$ 3.1\,arcsec of the transit source location for TOI-4504\,b (based on the S27-69 search).

\subsection{Follow-up photometry}
Several follow-up photometric observations of TOI-4504\,c were conducted and are available in Exo-FOP. Eight observations were scheduled to record the transit of TOI-4504\,c, assuming a constant period, but all resulted in non-detections. The non-detections, potentially due to the large TTVs, make it harder to predict the next transit. We include our predictions for upcoming transits in \autoref{sec:prediction}.
There are two observations of TOI-4504\,b, but since this exoplanet does not directly contribute to detected TTVs aimed to be studied in this work, this observation was excluded from the modeling scheme.

\begin{table}[ht!]
\caption{Priors and posteriors for TOI-4504\,b parameters derived with {\textsc juliet}.}
\centering
\label{02_transit_fit}
\begin{tabular}{ccc}
 \hline 
  \hline
Parameter       & Prior                         & Posterior                           \\ \hline
$P$ [days]               & $\mathcal{N}$ (2.4,0.5)       & $2.42614^{+0.00014}_{-0.00013}$           \\
$t_0$ [BJD]               & $\mathcal{N}$ (2459038.46,0.2)       & $2459038.458^{+0.022}_{-0.021}$           \\

$b$               & $\mathcal{U}$ (0.0,1.0)       & $0.396^{+0.134}_{-0.197}$           \\
$R_p/R_{\star}$               & $\mathcal{U}$ (0.0,1.0)       & $0.0268^{+0.0019}_{-0.0019}$           \\
$q_1$        & $\mathcal{U}$ (0.0,1.0)       & $0.527^{+0.320}_{-0.308}$           \\
$q_2$        & $\mathcal{U}$ (0.0,1.0)       & $0.496^{+0.323}_{-0.316}$           \\
$e$               & fixed 0.0                     & --                                  \\
$\omega$ [$^\circ$]           & fixed 90.0                    & --                                  \\
$\rho_{\star}$ [kg cm$^{-3}$]             & $\mathcal{N}$ (1600,100)      & $1601^{+100}_{-97}$                \\
$m_{\rm dilution}$ & fixed, 1.0                    & --                                  \\
$m_{\rm flux}$     & $\mathcal{N}$ (0.0,0.1)       & $-0.000046^{+0.000014}_{-0.000014}$ \\
$\sigma_w$ [ppm]  & $\mathcal{J}$ (0.1,1000.0)    & $2.16^{+17.11}_{-1.90}$                   \\
$a$ [au]               & --                            & $0.03392 \pm 0.00068$                  \\
$R_p$ [$R_{\oplus}$]   & --                            & $2.691 \pm 0.191$                  \\
$i$ [$^\circ$]   & --                            & $87.4^{+0.9}_{-1.3}$                  \\
\hline
\hline
\end{tabular}
\tablecomments{$a$ is calculated using Kepler's third law and derived period $P$}
\end{table}

We conclude that TOI-4504 is a metal-rich ([Fe/H]$ = 0.16\pm0.05$\,dex) main sequence star ($M_{\star} = 0.89^{+0.06}_{-0.04}$\,M$_{\odot}$, $R_{\star} = 0.92^{+0.04}_{-0.04}$\,R$_{\odot}$), just on the boundary between a G- and K-type star ($T_{\rm{eff}} = 5315 \pm 60$\,K).

\section{Stellar parameters of TOI-4504}
\label{sec3}

We followed the two-step iterative procedure presented in \citet{wine1} to obtain the stellar parameters of the host star. We start from the co-added FEROS spectra to obtain the stellar atmospheric parameters ($T_{\rm{eff}}$, $\log g$, [Fe/H], $v \sin i$) using the \textsc{zaspe} package \citep{zaspe}.
We perform a spectral energy distribution (SED) fit to estimate the stellar physical parameters, using the publicly available broad-band photometry as data and the PARSEC isochrones as a model. This step involved the use of the GAIA DR3 \citep{gaia_b} parallax to convert the observed apparent magnitudes to absolute magnitudes and adopt a simple interstellar extinction rule \citep{cardelli:1989}. 

We also fix the metallicity to the one derived with \textsc{zaspe} and use the \textsc{zaspe} value for the $T_{\rm{eff}}$ as a prior. This step produces a more precise value for the stellar $\log g$, which is used as input in a new run of \textsc{zaspe}. We continue with the iterations until reaching convergence of the $T_{\rm{eff}}$ and [Fe/H] in two subsequent \textsc{zaspe} runs.
Stellar parameters from our analysis are listed in \autoref{star_params}.
The uncertainties in the stellar parameters obtained with our procedure do not include possible systematic differences with respect to other stellar models; because of this, we inflate the uncertainties as suggested in \cite{Tayar_2022}.

\section{Analysis and Results}
\label{sec4}


\begin{figure}[ht]
  \centering
  \includegraphics[width=1\linewidth]{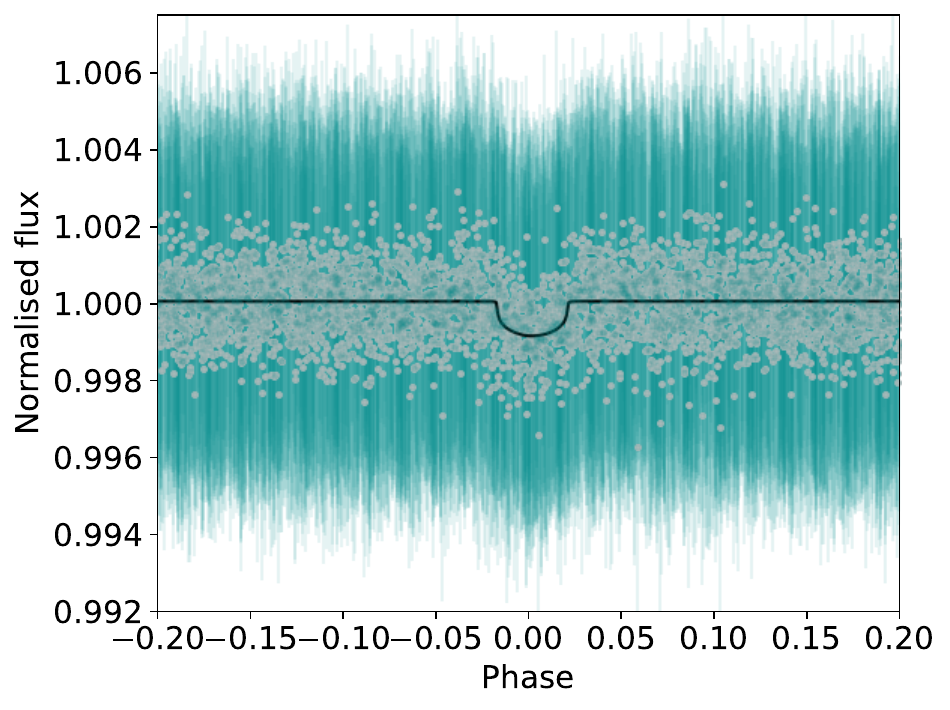}
  \caption{Phase plot for TOI-4504\,b transit. Light curve was binned into one-hour bins.}
  \label{02_transit_fit_fig}
\end{figure}

\subsection{Transit analysis and TTV extraction}
\label{sec:ttv}

For the transit fitting, we used the python package {\textsc juliet} \citep{juliet}, which employs transit light-curve models from the {\textsc batman} package \citep{batman}. Transit analysis of each planet was treated individually.

\subsubsection{TOI-4504\,b}
\label{sec:4504b}
For the transit analysis of TOI-4504\,b, we used 2-min PDCSAP data. Before the analysis, we deleted transits of TOI-4504\,c from the time series. We used broad uninformative priors (see \autoref{02_transit_fit}). The final values of the fit are also listed in \autoref{02_transit_fit}.

\autoref{02_transit_fit_fig} shows binned data that are phase-folded with the orbital period of 2.42614\,days together with the fit of the transit. We detected a planet with a transit depth of $\Delta F=718$\,ppm. 
Our RV data were insufficient to confirm this planet and estimate its mass (see \autoref{feros_anaysis}). However, the radius of $R_{\rm p}=2.691 \pm 0.191$\,R$_{\oplus}$ and mass of $10.4 \pm 0.9$\,M$_{\oplus}$ calculated from the mass-radius relations \citep{muller:2024} point towards the classification of planet b as sub-Neptune. We did not detect TTVs in TOI-4504\,b.

\subsubsection{TOI-4504\,c}
\label{sec:4504c}
To extract TOI-4504\,c TTVs, we again used {\textsc juliet}. First, we detrended the light curves of all sectors containing transits of TOI-4504\,c. We used the Gaussian Process (GP) on the out-of-transit data to remove trends in each TESS sector before fitting the planet parameters. The GP model utilizes approximate Matern kernel from {\textsc celerite} \citep{celerite}.

We used wide priors for an amplitude of the GP $\sigma_{\rm{GP,TESS}}$ of $\mathcal{J}(10^{-6},10^{6})$, where $\mathcal{J}(a,b)$ represents a Jeffreys prior between $a$ and $b$. For the time/length-scale of the GP $\rho_{\rm{GP,TESS}}$ we used $\mathcal{J}(10^{-3},10^{3})$. 
The fit was done using priors for the instrumental parameters, namely the flux offset $m_{\rm{flux}}$, jitter $\sigma_{\rm{w}}$, dilution factor $m_{\rm{dilution}}$ and limb-darkening coefficients $q_{\rm{1}}$ and $q_{\rm{2}}$. For the planetary parameters, we used parametrization using planet-to-star radius ratio $p$ and impact parameter $b$. Priors for eccentricity $e$ and argument of periastron $\omega$ were fixed to 0 and 90$^{\circ}$, respectively.

\begin{figure}
\centering
\includegraphics[width=1\linewidth]{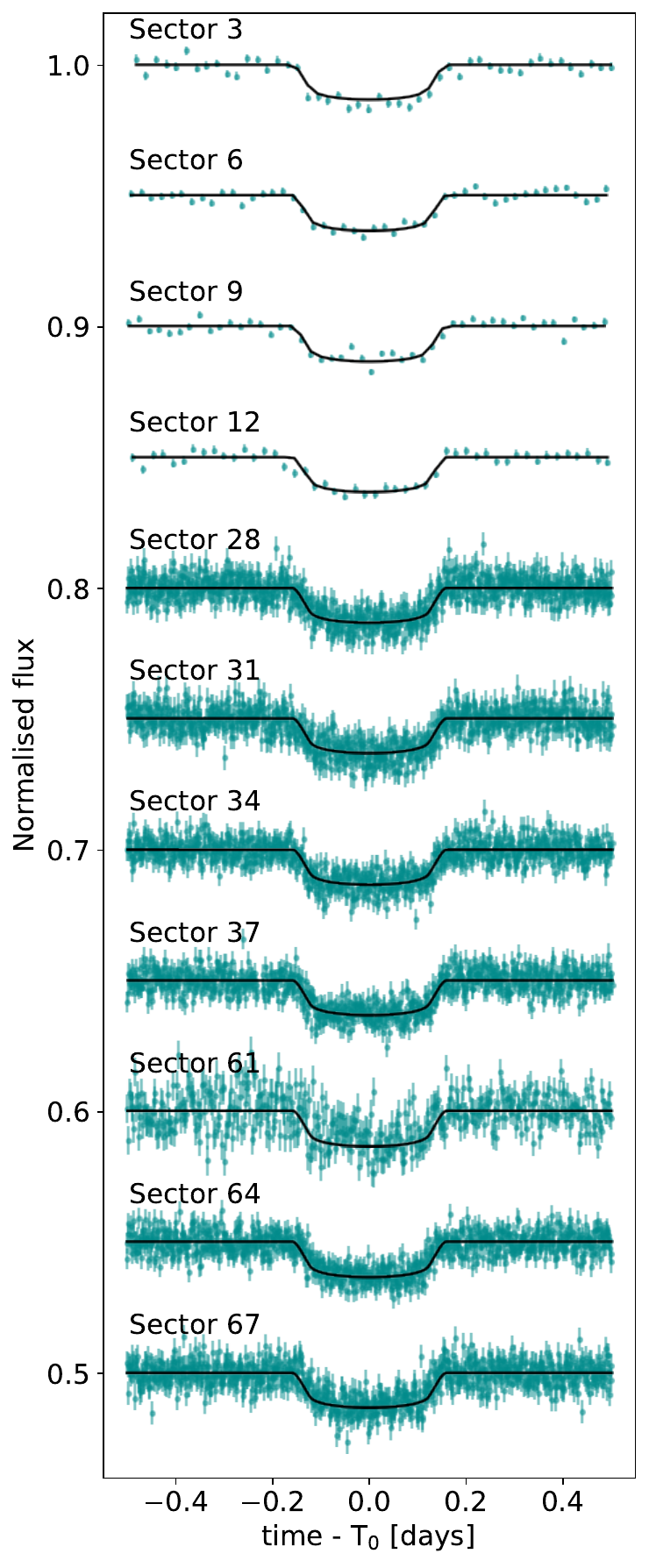}
\caption{Transits of TOI-4504\,c with a model from {\textsc juliet} shifted to have mid-transit at 0 and plotted with vertical offsets.}
\label{trans} 
\end{figure}

The RV data showed no indication of a significant eccentricity. Thus, fixed eccentricity is sufficient for TTVs determination. Priors for transit times $T_{\rm transit\,number}$ were determined visually from the light curve because of the strong TTVs. All priors and posteriors for the complete fit are listed in \autoref{ttv_juliet}. A zoom-in for the fit of the transits is shown in \autoref{trans}. 

\subsection{Spectroscopic data analysis}
\label{feros_anaysis}

\begin{figure}
  \centering
  \includegraphics[width=1\linewidth]{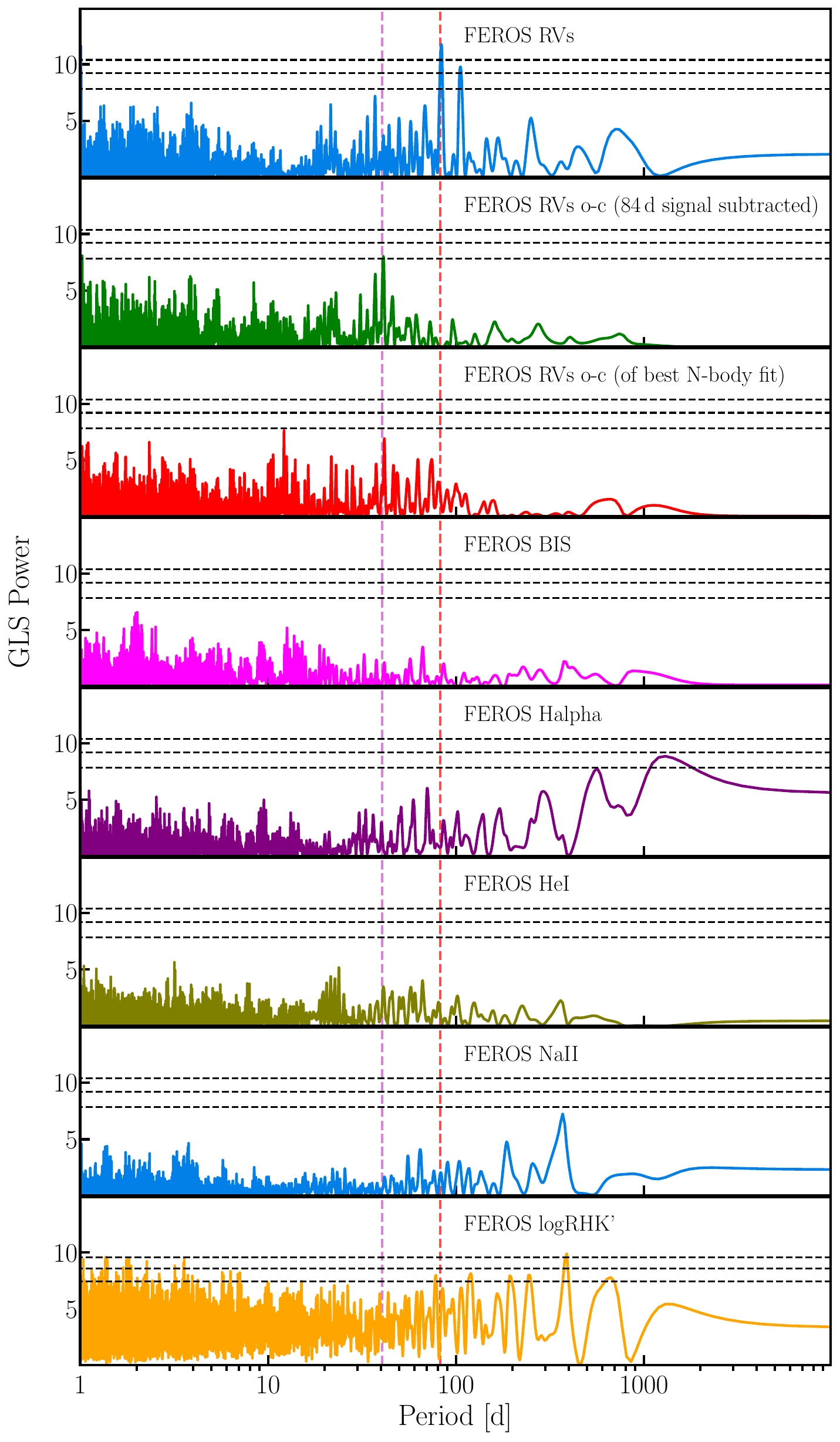}
  \caption{GLS power spectrum of FEROS spectroscopic products of TOI-4504. From top to bottom panels, as labeled, RVs used in this work, RV residuals after subtracting the dominant signal of TOI-4504\,c at 84\,d, the final the best-fit TTV+RV model residuals, BIS, $\rm{H}_{\alpha}$, He I  Na II, and $\log(R'_{HK})$ activity indicators, respectively. False alarm probability levels of 10\%, 1\%, and 0.1\% are marked with dashed lines, respectively. The red and magenta vertical lines indicate the best-fit periods of the transiting Jovian planet TOI-4504\,c, and the non-transiting TOI-4504\,d, respectively. }
  \label{activity}
\end{figure}

We performed a period search in the FEROS RVs and activity indices data using a generalized Lomb-Scargle periodogram \citep[GLS;][]{Zechmeister2009}. \autoref{activity} shows a stacked GLS periodogram of the FEROS spectroscopic data of TOI-4504. We found significant RV periodicity with a period of 84\,d, consistent with the detected quasi-periodic signal of TOI-4504\,c. The semi-amplitude of the 
84\,d signal in the FEROS data is $\sim$185\,m\,s$^{-1}$, corresponding to a planetary mass of $\sim$3.5\,$M_{\rm Jup}$, validating the planetary nature of TOI-4504\,c.

After subtracting this period, no other significant signals were detected in the RV residuals, although we recorded a very large RV scatter of the order of $\sim$100\,m\,s$^{-1}$. Keeping in mind the relatively low number of RV data and the prior knowledge of the complexity of the TOI-4504 system having at least three planets, two of them Jovian-mass, and the third a short-period Neptune, the RV jitter could be attributed to unresolved planet signals.

The second panel of \autoref{activity} reveals a prominent peak at a period of 41.2\,d, crossing the 10\% FAP threshold. As we will demonstrate in our results, 
this signal is likely induced by the non-transiting giant TOI-4504\,d. Moreover, 
subtracting this signal by simultaneously modeling it alongside the dominant 84\,d 
period failed to account for the observed large RV scatter. As shown in 
\autoref{nbody_models}, even a more sophisticated N-body modeling approach could not fully resolve the observed RV scatter.
A GLS analysis of the N-body model RV residuals (see third panel of \autoref{activity}) shows several peaks with periods of 2.33\,d, 12.16\,d, and 41.66\,d; the latter is close to the osculating period of TOI-4504\,d. However, all peaks are insignificant and likely emerged by chance, unrelated to planetary signals.
We concluded that the precision of the RV data was sufficient to constrain the mass of TOI-4504\,c, and when constrained by the TOI-4504\,c, the TTV signal could constrain the period and mass of the non-transiting massive 
planet. However, the FEROS RVs were insufficient for constraining parameters of the hot planet TOI-4504\,b, which has an expected RV semi-amplitude of only a few m\,s$^{-1}$.

The BIS, $\rm{H}_{\alpha}$, He\,I, and Na\,II did not show significant periodicity. However, it is worth mentioning that this is not the case for the $\log(R'_{HK})$ activity indicator, which showed many marginally significant periods, indicating that TOI-4504 is an active star, which may partially explain the large RV scatter. The large RV scatter in the FEROS data likely results from stellar activity, but we do not rule out additional planets in the system, an analysis of which is beyond the scope of this work.

\begin{table}[]
\caption{Priors and posteriors for the TTV extraction with {\textsc juliet} for TOI-4504\,c.}
\centering
\label{ttv_juliet}
\begin{tabular}{ccc}
 \hline 
  \hline
Parameter       & Prior                         & Posterior                           \\ \hline
$b$               & $\mathcal{U}$ (0.0,1.0)       & $0.492^{+0.038}_{-0.041}$           \\
$R_p/R_{\star}$               & $\mathcal{U}$ (0.0,1.0)       & $0.108^{+0.001}_{-0.001}$           \\
$q_1$        & $\mathcal{U}$ (0.0,1.0)       & $0.426^{+0.114}_{-0.125}$           \\
$q_2$        & $\mathcal{U}$ (0.0,1.0)       & $0.203^{+0.094}_{-0.091}$           \\
$e$               & fixed 0.0                     & --                                  \\
$\omega$ [$^\circ$]           & fixed 90.0                    & --                                  \\
$\rho_{\star}$ [kg cm$^{-3}$]             & $\mathcal{N}$ (1600,100)      & $1567^{+101}_{-100}$                \\
$m_{\rm dilution}$ & fixed, 1.0                    & --                                  \\
$m_{\rm flux}$     & $\mathcal{N}$ (0.0,0.1)       & $-0.000002^{+0.000012}_{-0.000012}$ \\
$\sigma_w$ [ppm]  & $\mathcal{J}$ (0.1,1000.0)    & $949^{+22}_{-25}$                   \\
$T_0$ [BJD] & $\mathcal{N}$ (2458401.4,0.5) & $2458401.4086^{+0.0032}_{-0.0032}$  \\
$T_1$ [BJD]  & $\mathcal{N}$ (2458483.2,0.5) & $2458483.2110^{+0.0034}_{-0.0034}$  \\
$T_2$ [BJD] & $\mathcal{N}$ (2458565.1,0.5) & $2458565.0902^{+0.0032}_{-0.0033}$  \\
$T_3$ [BJD] & $\mathcal{N}$ (2458647.3,0.5) & $2458647.3328^{+0.0043}_{-0.0045}$  \\
$T_8$ [BJD] & $\mathcal{N}$ (2459065.2,0.5) & $2459065.2370^{+0.0024}_{-0.0023}$  \\
$T_9$ [BJD] & $\mathcal{N}$ (2459148.5,0.5) & $2459148.4782^{+0.0027}_{-0.0026}$  \\
$T_{10}$ [BJD] & $\mathcal{N}$ (2459231.1,0.5) & $2459231.1144^{+0.0021}_{-0.0021}$  \\
$T_{11}$ [BJD]& $\mathcal{N}$ (2459313.3,0.5) & $2459313.2535^{+0.0019}_{-0.0019}$  \\
$T_{19}$ [BJD]& $\mathcal{N}$ (2459976.1,0.5) & $2459976.0493^{+0.0043}_{-0.0045}$  \\
$T_{20}$ [BJD]& $\mathcal{N}$ (2460059.6,0.5) & $2460059.6189^{+0.0023}_{-0.0023}$  \\
$T_{21}$ [BJD]& $\mathcal{N}$ (2460142.6,0.5) & $2460142.6048^{+0.0022}_{-0.0023}$  \\
$P$ [days]               & --                            & $82.97213^{+0.00013}_{-0.00013}$    \\
$t_0$  [BJD]            & --                            & $2458400.3906^{+0.0016}_{-0.0016}$  \\
$a$ [au]               & --                            & $0.3546^{+0.0073}_{-0.0077}$                  \\
$R_p$ [$R_{\rm Jup.}$]   & --                            & $0.99 \pm 0.05$                  \\
$i$ [$^\circ$]   & --                            & $89.69^{+0.02}_{-0.03}$                  \\
\hline
\hline
\end{tabular}
\tablecomments{$P$ was computed as a slope and $t_0$ as an intercept of a least-squares fit to the transit times}
\end{table}

\subsection{Orbital modeling based on TTV and RV data}
\label{nbody_models}


In this section, we report the orbital analysis results of TOI-4504\,c and its indirectly detected neighbour, which, from now on, we name TOI-4504\,d. We analyzed TOI-4504\,c \& d separately because, as we discussed in \autoref{feros_anaysis}, we were not able to set constraints on the mass of the transiting planet TOI-4504\,b with the FEROS RV data, nor with detailed three-planet N-body modeling of the TESS transits and TTVs 
separately or together with the RV data. We note in passing that we performed a three-planet orbital analysis including TOI-4504\,b, but including TOI-4504\,b did not significantly improve the fit compared to the models considering only TOI-4504\,c \& d, and thus shall not be discussed in this work.

We conducted a joint global orbital analysis using broad priors to explore the parameter space consistent with the TTV and RV data for TOI-4504, using the {\textsc Exo-Striker}\footnote{Available at \url{https://github.com/3fon3fonov/exostriker}} exoplanet modeling toolbox \citep{exostriker}. Our fitting scheme followed a more targeted N-body orbital fit once the consistent parameter space had been identified.
Taking the gravitational 
interactions between TOI-4504\,c \& d into account, the {\textsc Exo-Striker} uses an internal RV dynamical model, whereas the TTV model is wrapped around the \mbox{\textsc TTVfast} package \citep{Deck2014}. We follow a very similar TTV+RV N-body fitting approach, used by \citet{TOI-2202} for the TOI-2202 system, which is similar to TOI-4504 in many aspects. We refer the reader to \citet{TOI-2202} for more details, and below we summarize our fitting methods, used parameters, adopted priors, and more results relevant to our modeling cascade, which reveal the orbital and physical parameters of the TOI-4504\,c \& d pair.

The fitted parameters for each planet in the TTV+RV model were the RV semi-amplitude $K_{\rm c,d}$ and the osculating planetary orbital period $P_{\rm c,d}$. The eccentricities $e_{\rm c,d}$, arguments of periastron $\omega_{\rm c,d}$, and mean anomalies $M_{\rm c,d}$ were derived using the parameterization $h = e\sin\omega$, $k = e\cos\omega$, and $\lambda = \omega + M$, which is more efficient for orbits with small eccentricities \citep{Tan2013}. 
Since we know that the perturber planet is not transiting, we  assumed a non-coplanar, mutually inclined orbital geometry and fitted for the orbital inclinations $i_{\rm c,d}$ and the difference between the longitudes of the ascending nodes $\Delta\Omega_{\rm d - c}$ = $\Omega_{\rm d}$ - $\Omega_{\rm c}$, where the mutual inclination comes following the expression:
\begin{equation}
\Delta i = \arccos[\cos(i_d)\cos(i_{c})+\sin(i_d)\sin(i_{c})\cos(\Delta\Omega)] .
\label{eq:deltai}
\end{equation}

\noindent
Since only $\Delta\Omega$ is important, we kept $\Omega_{\rm d}$ fixed to 0$^{\circ}$, and we fit for $\Omega_{\rm c}$.

All osculating parameters in our N-body fitting were defined in the Jacobi coordinate system, which is standard practice for multiple-planet systems \citep[][]{Lee2003}, and are valid for the reference epoch BJD = 2458400.0, arbitrarily chosen before the first {\it TESS} transit of TOI-4504\,c.
Additionally, for the FEROS RV data, we fitted the RV offset and RV jitter parameters \citep[see,][]{Baluev2009}. For the stellar mass of TOI-4504, we used 0.885 $M_{\odot}$, which, together with the fitted parameters, was converted to dynamical planetary masses $m_{\rm c,d}$ needed to construct the N-body model. The numerical time step in the dynamical model was set to 0.5 days, which is sufficiently small for accurate modeling of the system.

\begin{figure*}
  \centering
  \includegraphics[width=1\linewidth]{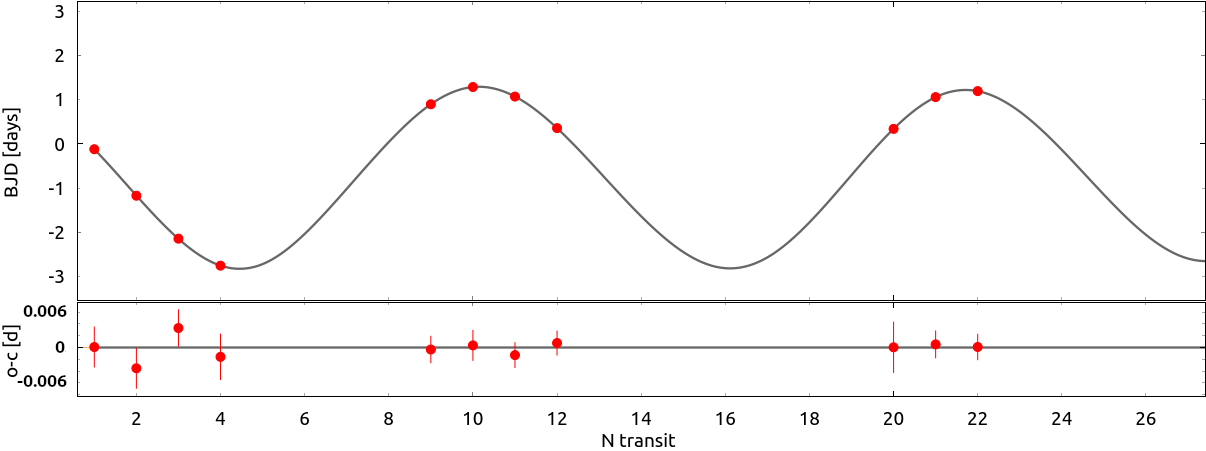}
    \includegraphics[width=1\linewidth]{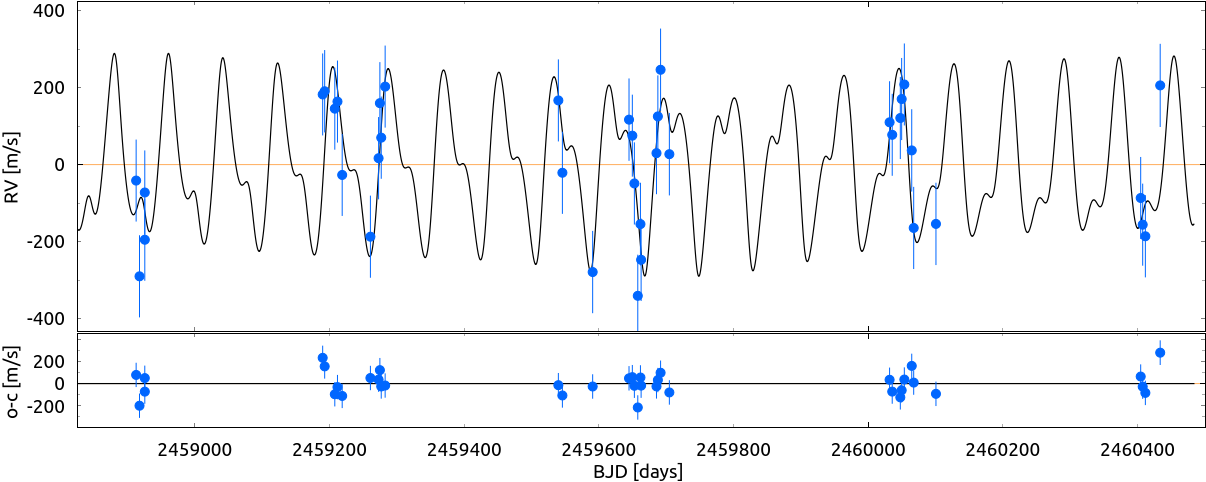}

  \caption{TESS TTV time series of TOI-4504\,c and a model consistent with two Jovian-mass planets with periods close to the 2:1 MMR commensurability, with the non-transiting planet being interior (top panel). The TTV signal is expressed as the deviation of the {\it TESS} transit events with respect to the mean osculating Period of TOI-4504\,c, which have a large semi-amplitude of $\sim$ 2\,d and super-period of 946.5\,d. The bottom sub-panel shows the TTVs residuals. The main bottom panel shows the Doppler component of the same model fitted to the FEROS RV data. The bottom sub-panel shows the RV residuals.}
  \label{TTVs}
\end{figure*}

    \begin{figure}
\includegraphics[width=0.49\linewidth]{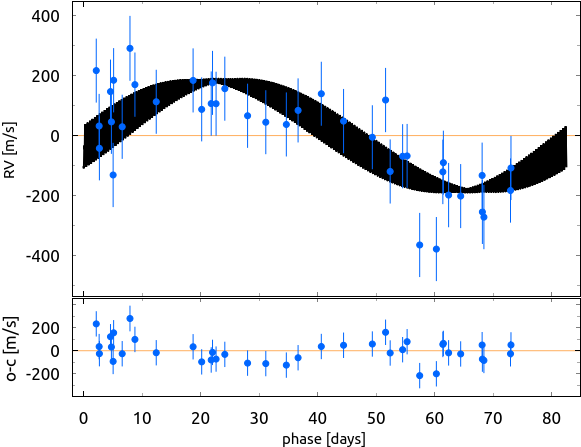}\put(-45,80){\scriptsize TOI-4504\,c }
\includegraphics[width=0.49\linewidth]{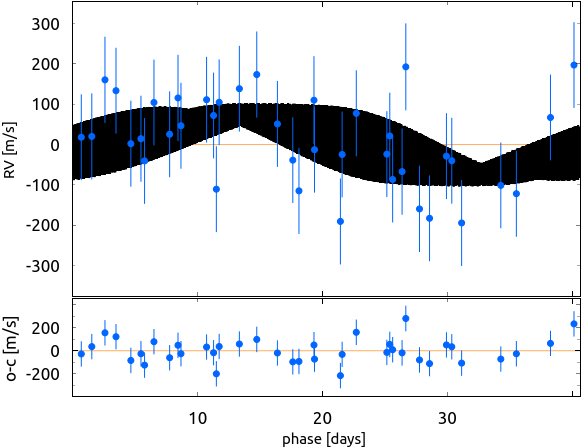}\put(-45,80){\scriptsize TOI-4504\,d }\\
\includegraphics[width=0.49\linewidth]{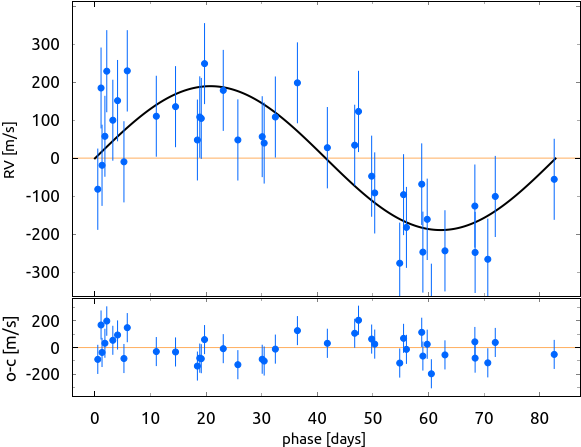}\put(-45,80){\scriptsize TOI-4504\,c }
\includegraphics[width=0.49\linewidth]{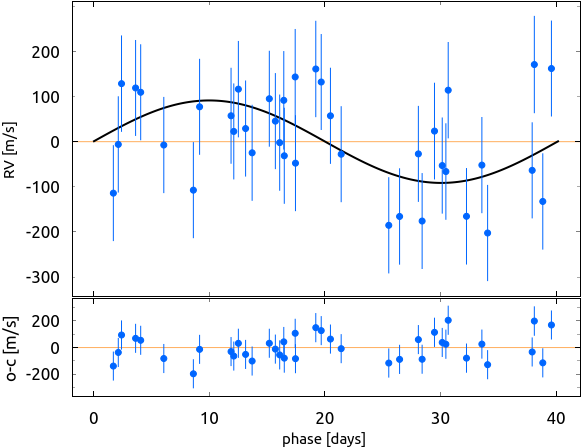}\put(-45,80){\scriptsize TOI-4504\,d }

\caption{Phased RV signals for the planets TOI-4504\,c and d. The top two 
panels display the planetary signals along with the osculating N-body model, phased 
to the best-fit periods from \autoref{NS_params}, valid for BJD = 2458400.0. The 
bottom two panels illustrate the same planetary signals, phased to the mean osculation periods derived from the TTV+RV model.}
\label{phased} 
\end{figure}

\begin{table*}[ht!]
\caption{{MCMC sampling priors, posteriors, and the optimum $-\ln\mathcal{L}$ orbital parameters of the two-planet system derived by joint N-body modeling of TTVs (TESS) and RVs (FEROS).}}
\label{NS_params}
\centering
 \begin{adjustwidth}{-2cm}{}
 \resizebox{0.83\textheight}{!}
 {\begin{minipage}{1.1\textwidth}

\begin{tabular}{lrrrrrrrrrrrr}     

\hline\hline  \noalign{\vskip 0.7mm}

\makebox[0.1\textwidth][l]{\hspace{45 mm} Median and $1\sigma$  \hspace{15 mm} Max. $-\ln\mathcal{L}$     \hspace{30 mm} Adopted priors  \hspace{10 mm} \hspace{1.5 mm} } \\
\cline{1-9}\noalign{\vskip 0.7mm}

Parameter &\hspace{10.0 mm} Planet d & Planet c &  & Planet d & Planet c  & & \hspace{10.0 mm}Planet d & Planet c  \\
\cline{1-9}\noalign{\vskip 0.7mm}   
 
$K$  [m\,s$^{-1}$]            & 90.8366$_{-2.5466}^{+1.8812}$ & 190.8921$_{-6.2119}^{+4.7269}$ &  
                              & 91.3360 & 189.2623 &
                              & $\mathcal{U}$(80.0, 140.0) & $\mathcal{U}$(150.0, 250.0)   \\ \noalign{\vskip 0.9mm}

$P$  [days]                    & 40.5634$_{-0.0368}^{+0.0363}$ & 82.5438$_{-0.0176}^{+0.0150}$ & 
                              & 40.5586 & 82.5383 &  
                              & $\mathcal{U}$(40.2, 40.6) & $\mathcal{U}$(81.0, 83.0)   \\ \noalign{\vskip 0.9mm}
                              
$e \sin(\omega)$               & 0.0439$_{-0.0011}^{+0.0010}$ & -0.0320$_{-0.0016}^{+0.0014}$ &
                              & 0.0441 & -0.0320 &  
                              & $\mathcal{U}$( -0.1, 0.1) & $\mathcal{U}$( -0.1, 0.1)   \\ \noalign{\vskip 0.9mm}

$e \cos(\omega)$               & -0.0064$_{-0.0047}^{+0.0039}$ & 0.0005$_{-0.0013}^{+0.0011}$ &
                              & -0.0027 & -0.0009 &  
                              & $\mathcal{U}$( -0.1, 0.1) & $\mathcal{U}$( -0.1, 0.1)   \\ \noalign{\vskip 0.9mm}

$\lambda$  [deg]               & 9.89$_{-2.43}^{+3.45}$  & 83.97$_{-0.15}^{+0.12}$ &
                              & 14.11 & 83.80 &   
                              & $\mathcal{U}$(0.0, 360.0) & $\mathcal{U}$(70.0, 110.0)  \\ \noalign{\vskip 0.9mm}

$i$  [deg]               & 85.00$_{-0.30}^{+0.28}$  & 89.69$_{-0.03}^{+0.03}$ &
                              & 84.74 & 89.68 &   
                              & $\mathcal{N}$(      86.0,    3.0)& $\mathcal{N}$(      89.7,    0.1)\\ \noalign{\vskip 0.9mm}

$\Delta\Omega$  [deg]               & $\dots$  & 0.0$_{-1.0}^{+0.9}$ &
                              & $\dots$ & -0.8&   
                              & $\dots$ & $\mathcal{N}$(0.0, 30.0)  \\ \noalign{\vskip 0.9mm}

RV$_{\rm off.}$ FEROS  [m\,s$^{-1}$]  & $\dots$  & 2067.0517$_{-14.8783}^{+14.2161}$  &     & $\dots$ &  2064.9642    &      & $\dots$ & $\mathcal{U}$(1900.0, 2200.0)  \\ \noalign{\vskip 0.9mm}
RV$_{\rm jit.}$ FEROS  [m\,s$^{-1}$]  & $\dots$ & 103.3721$_{-7.0042}^{+13.8367}$   &  &  $\dots$ & 104.6664     &         & $\dots$ & $\mathcal{U}$(1.0, 150.0)  \\ \noalign{\vskip 0.9mm}


$e$                             & 0.0445$_{-0.0009}^{+0.0010}$ & 0.0320$_{-0.0014}^{+0.0016}$ &
                              & 0.0441 & 0.0320 &  
                              & (derived)  & (derived)   \\ \noalign{\vskip 0.9mm}

$\omega$ [deg]                & 98.3$_{-5.1}^{+6.1}$ & 270.9$_{-2.2}^{+2.0}$ &
                              & 93.5 & 268.4&  
                              & (derived)  & (derived)   \\ \noalign{\vskip 0.9mm}

$M_0$  [deg]                 & 271.6$_{-7.5}^{+7.3}$  & 173.1$_{-1.9}^{+2.1}$ &
                              & 280.6 & 175.4 &   
                              & (derived)  & (derived)   \\ \noalign{\vskip 0.9mm}

$\Delta i$  [deg]             & $\dots$ & 4.7$_{-0.3}^{+0.3}$   &  
                              &   $\dots$ & 5.0 &    
                              &     $\dots$  & (derived) & \\ \noalign{\vskip 0.9mm}

$m_p$  [M$_{\rm Jup.}$]          & 1.4166$_{-0.0647}^{+0.0651}$  & 3.7672$_{-0.1822}^{+0.1810}$ & 
                              & 1.4294 & 3.7494 &     
                              & (derived) & (derived)   \\ \noalign{\vskip 0.9mm}

$a_p$  [au]                     & 0.2219$_{-0.0043}^{+0.0041}$  & 0.3569$_{-0.0069}^{+0.0066}$  &  
                              & 0.2219 & 0.3569 &     
                              & (derived) & (derived)   \\ \noalign{\vskip 0.9mm}
 
\hline \hline \noalign{\vskip 0.7mm}

\end{tabular}
\end{minipage}}
\end{adjustwidth}
\tablecomments{The orbital elements are in the Jacobi frame and are valid for epoch BJD = 2458400.0. The adopted priors are listed in the right-most columns, and their meanings are $\mathcal{U}$ -- Uniform, and $\mathcal{N}$ -- Gaussian priors.
The derived planetary posterior parameters of $a$, and $m$ are calculated by considering the stellar parameter uncertainties.}
\end{table*}

\begin{figure*}

\includegraphics[width=9cm]{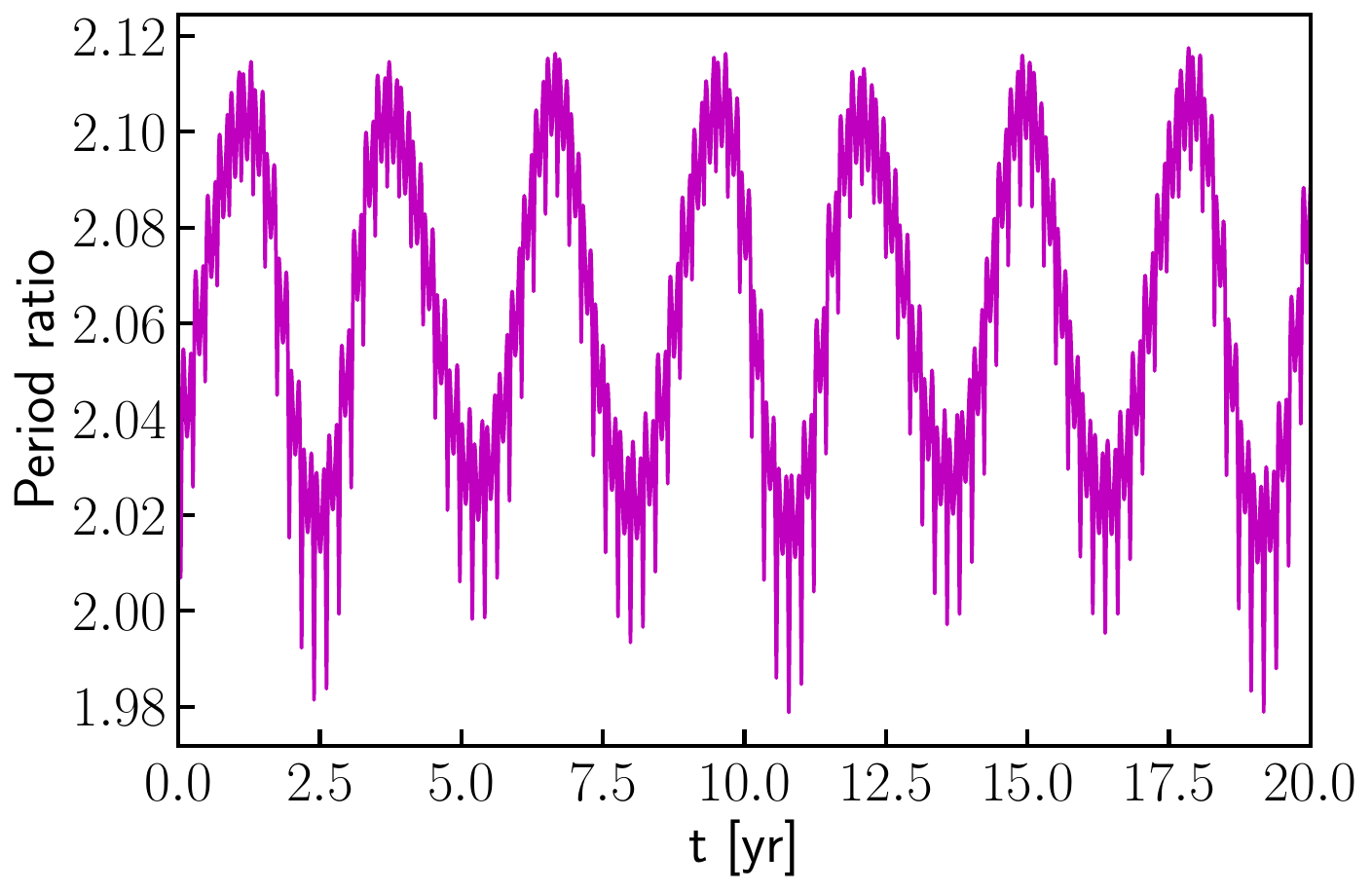}
\includegraphics[width=9cm]{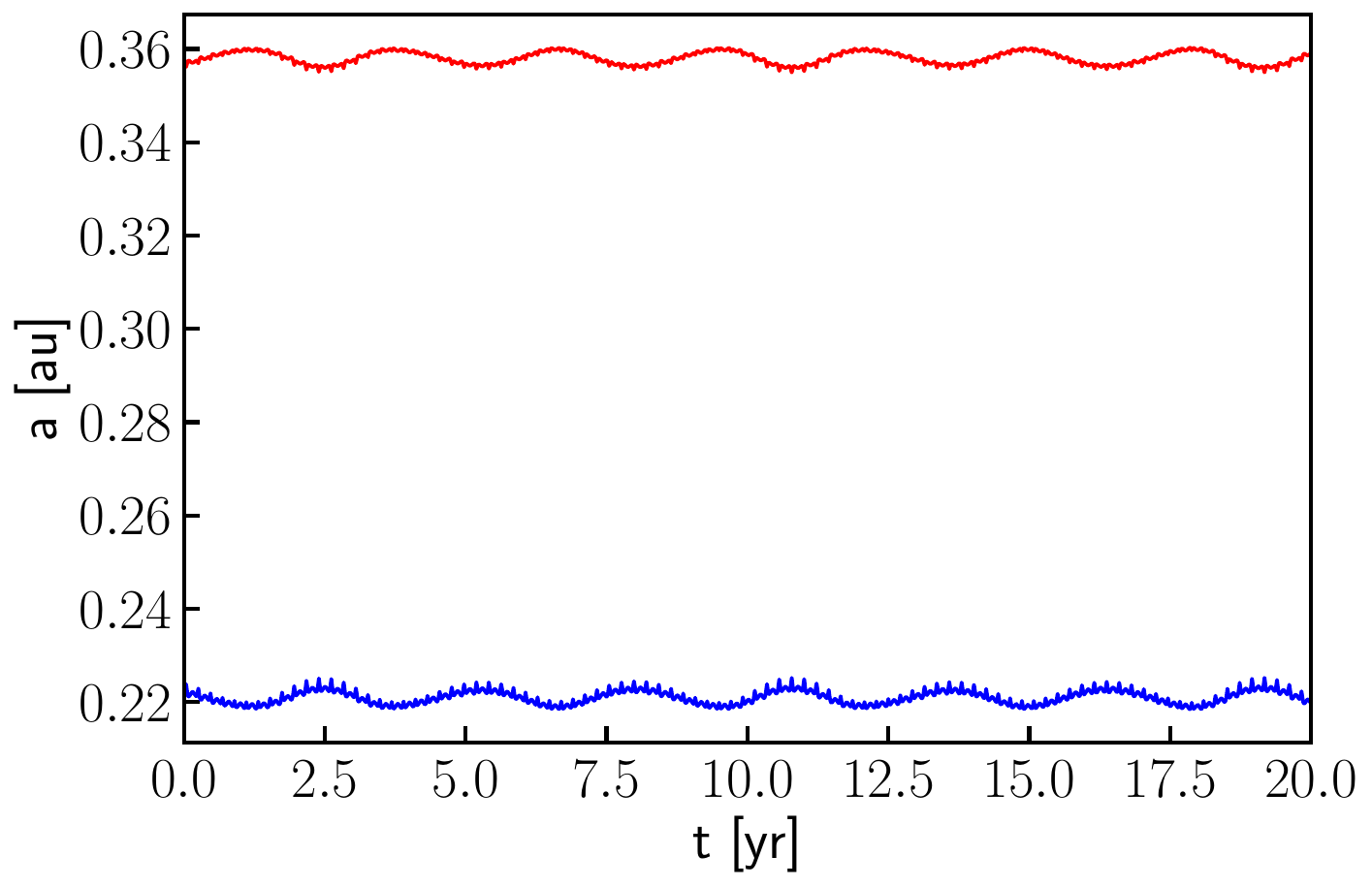} \\
\includegraphics[width=9cm]{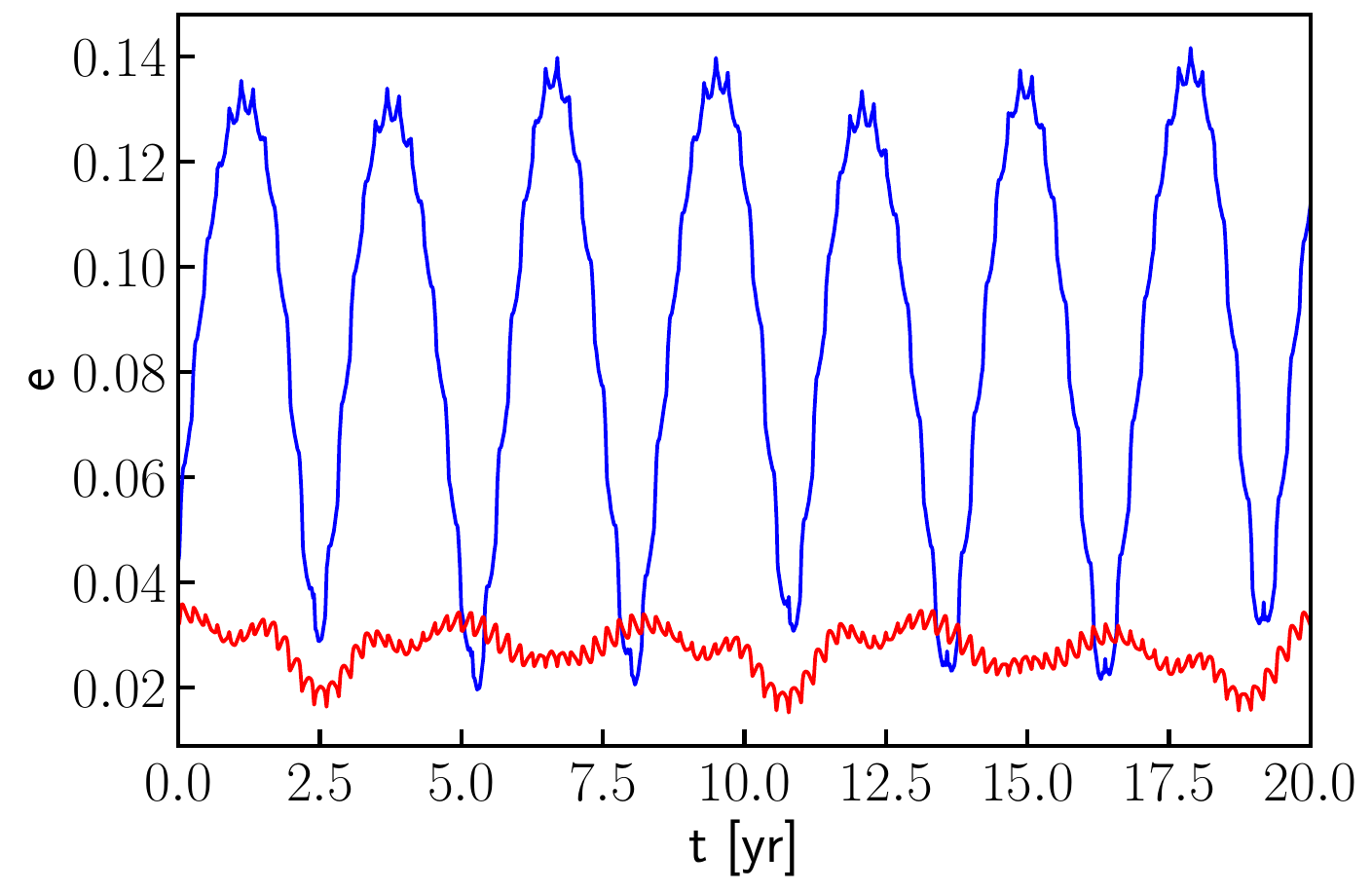} 
\includegraphics[width=9cm]{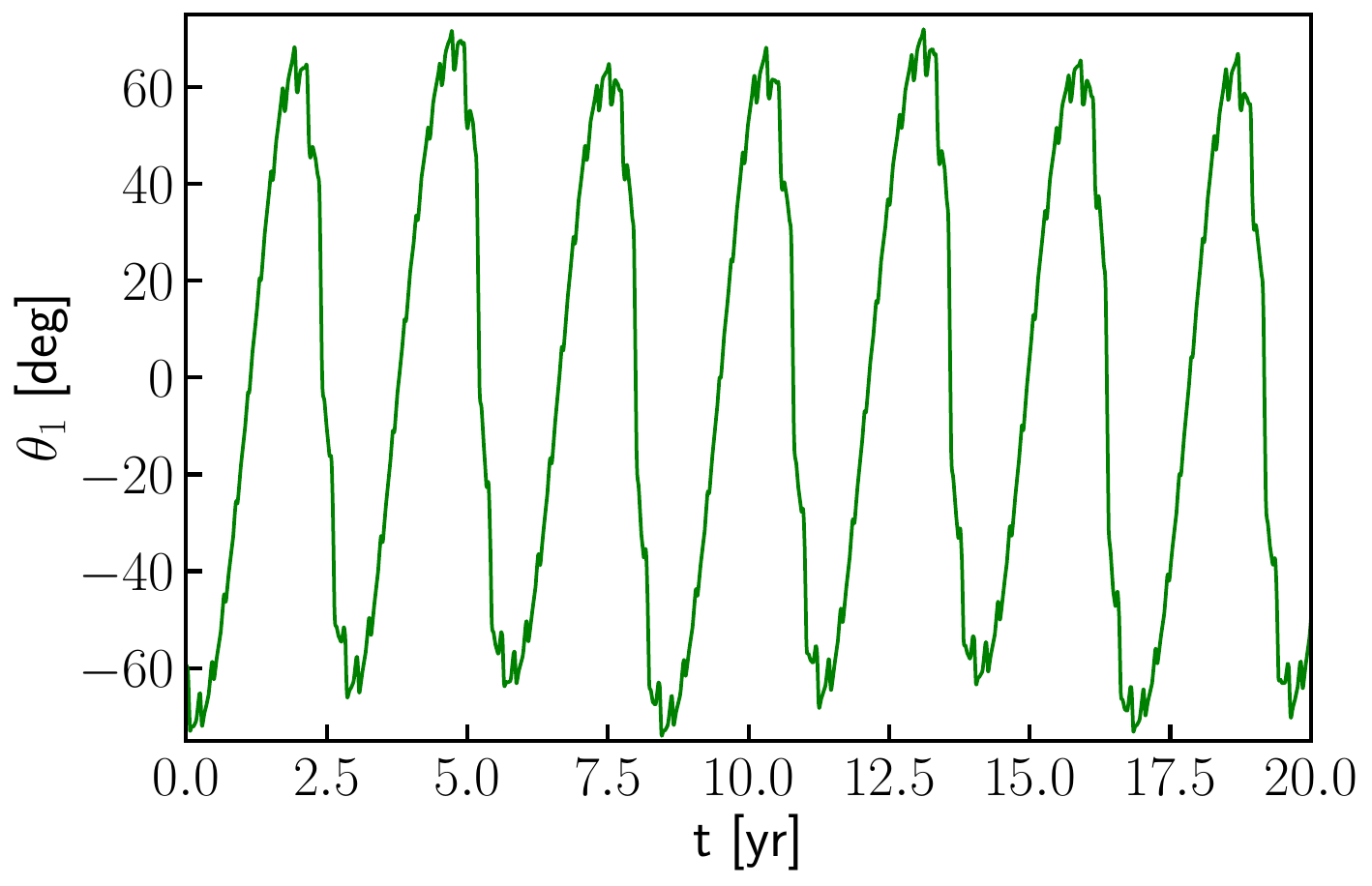} 
\caption{Orbital evolution of the best TTV+RV N-body model of the TOI-4504 system for a short extent of 20\,yr long N-body simulation using the {\textsc Exo-Striker}. The top row from left to right panels shows the evolution of the planetary period ratio ($P_{\rm c} / P_{\rm d}$) (magenta) and the evolution of the semi-major axes $a_{\rm c}$ (red) and $a_{\rm d}$ (blue), respectively. The bottom row from left to right panels shows the evolution of the eccentricities $e_{\rm c}$ (red) and $e_{\rm d}$ (blue) and $\theta_1$ (green), which librates around 0$^{\circ}$ with amplitude of $\sim$ 65$^{\circ}$, respectively. }
\label{evol_plot} 
\end{figure*}

We first used a global Nested Sampling (NS) parameter search \citep{Skilling2004} via the {\textsc dynesty} sampler \citep{Speagle2020}. We employed a Static-NS setup with 500 "live points" per fitted parameter \citep[see][for details]{Speagle2020}, to explore a wide parameter space of 
eccentricities, masses, and periods for the "yet-to-be detected" planet TOI-4504\,d, assuming it could be either interior or exterior to TOI-4504\,c. For TOI-4504\,c, the prior parameter range estimates were taken from our GLS and TTV extraction analysis, making the parameter search more constrained. The adopted prior ranges for TOI-4504\,d \& c are listed in \autoref{globalNS}. Our results showed very poor fits when TOI-4504\,d was assumed to be exterior. In contrast, good fits were found in the resulting posterior probability distribution analysis when the planet was considered interior. As indicated in \autoref{Nest_samp_ttv}, the posteriors are multimodal but firmly converge with TOI-4504\,d being an interior planet in the 2:1 period ratio commensurability.

As a next step, we used the NS posteriors and the best $-\ln\mathcal{L}$ NS solution as an initial guess for a more targeted Simplex optimization \citep{NelderMead}, aiming to find the optimal $-\ln\mathcal{L}$ best-fit solution. This was followed by an affine-invariant ensemble Markov Chain Monte Carlo (MCMC) sampler \citep{Goodman2010} via the \texttt{emcee} package \citep{emcee} to build the parameter posteriors and obtain parameter uncertainties.

\autoref{TTVs} shows our best joint fit to the TTVs and RVs, whereas 
\autoref{mcmc_posteriors} displays the MCMC posteriors drawn around the best fit. \autoref{NS_params} lists the best fit in terms of optimal $-\ln\mathcal{L}$ shown in \autoref{TTVs}, the median orbital parameters and their uncertainties extracted from our MCMC posteriors, and the used priors.

\autoref{phased} shows the phased RV planet signals of TOI-4504\,c and d, respectively.
The top two panels of \autoref{phased} show the phased planetary periods to the N-body fit, which, due to the dynamical nature of the model are strong osculating. In contrast, the bottom two panels provide a clearer representation of the RV signals, phased to the main osculation periods from the TTV+RV N-body model, which we estimate to be 82.5438$_{-0.0176}^{+0.0150}$\,d, and 40.5634$_{-0.0368}^{+0.0363}$\,d for TOI-4504\,c and d, respectively.
We conclude that the TTVs from TESS and the RV data from FEROS suggest the presence of a massive pair of Jovian planets with osculating periods of $P_{\rm d}$ = 40.5634$_{-0.0368}^{+0.0365}$\,days and $P_{\rm c}$ = 82.5438$_{-0.0150}^{+0.0176}$\,days, and small but significantly non-zero eccentricities of $e_{\rm d}$ = 0.0445$_{-0.0009}^{+0.0010}$ and $e_{\rm c}$ = 0.0320$_{-0.0014}^{+0.0016}$, valid for reference epoch of BJD = 2458400.0. We obtain dynamical masses of $m_{\rm d}$ = 1.417$_{-0.065}^{+0.065}$\,M$_{\rm Jup}$ and $m_{\rm c}$ = 3.767$_{-0.181}^{+0.182}$\,M$_{\rm Jup}$ for TOI-4504\,d and TOI-4504\,c, respectively. The mutual inclination between the two planets suggests a non-coplanar configuration with $\Delta i$ = 4.7$_{-0.3}^{+0.3}$ degrees, which agrees with the fact that TOI-4504\,d is not transiting.

\subsection{Dynamics and long-term stability}
\label{sec4.3}


We performed N-body simulations to study the long-term stability 
and dynamical evolution of the TOI-4504 system. For this test, 
we again ignored the innermost planet TOI-4504\,b, since, even if its 
dynamical mass is in the Neptune mass regime, its overall mutual 
Hill distance with TOI-4504\,d would be $\sim$ 17 R$_{\rm Hill,m}$ \citep[see,][]{Gladman1993}, thus would result in minimal dynamical interactions with the outer pair. 
Additionally, including TOI-4504\,b would require using a very 
small time step of approximately 30 minutes (1/100 of the orbital period of the planet) to achieve accurate orbital dynamics, which would be computationally expensive.

Our study of the long-term dynamics and possible MMR dynamics in the system adopted the same N-body numerical setup used in our recent analyses of the TOI-2202 \citep{TOI-2202} and TOI-2525 \citep{Trifonov2023} systems, which share similar 
physical and orbital characteristics with the Jovian pair of 
TOI-4504. We conducted numerical integrations using a custom 
version of the Wisdom-Holman N-body integrator \citep{Wisdom1991}, 
specifically tailored for long-term stability analyses of exoplanetary systems in Jacobi orbital elements. Integrated natively within the {\sc Exo-Striker} toolbox, this N-body integrator allows direct injection of posterior samples in Jacobi coordinates, thereby avoiding the additional coordinate transformations typically required by other publicly available N-body integrators.
We 
tested the stability of the TOI-4504 system for a maximum of 10 million years 
with a time step of 0.5 days for 10,000 randomly chosen samples 
from the achieved orbital parameters of the TTV+RV MCMC parameter 
posteriors. For each sample integration and at each numerical time step, we monitor the planetary semi-major axes, ensuring they do 
not deviate by more than 20\% from their initial values; any orbit 
exceeding this threshold is considered unstable, resulting in the 
termination of the integration. Additionally, integrations are flagged 
as unstable and terminated if eccentricity values become sufficiently 
excited to orbit-crossing configurations. Further details on our 
stability criteria are provided in \citet{TOI-2202}.

We found that all examined parameters are long-term stable. From the numerical evolution, we built a posterior distribution of some of the important dynamical properties of the system, such as the mean period ratio $P_{\rm rat.}$, mean eccentricities $\hat e_{\rm d}$, $\hat e_{\rm c}$, their peak-to-peak amplitudes $e_{\rm d}$, $e_{\rm c}$, the dynamical masses of the planets $m_{\rm d}$, $m_{\rm c}$, and the orbital semi-major axes $a_{\rm d}$, $a_{\rm c}$, as in \citet{wine5,Trifonov2023}. We also collected posterior distributions of the resonance angles defined as:

\begin{equation}
\Delta\omega = \omega_{\rm c} - \omega_{\rm d}
\label{eq1}
\end{equation}

\noindent
which is the secular apsidal angle, and:

\begin{equation}
\theta_1 =  \lambda_{\rm c} - 2\lambda_{\rm d} + \omega_{\rm c},~~~~~~\theta_2 =  \lambda_{\rm c} - 2\lambda_{\rm d} + \omega_{\rm d}
\label{eq2}
\end{equation}

\noindent
are the first-order eccentricity-type 2:1 MMR angles, of which at least one must librate 
around a fixed angle to claim the system in an MMR \citep[see][]{Lee2004}. 

We find that the TOI-4504 system exhibits moderate eccentricity evolution, with the less massive planet, TOI-4504\,d, showing larger eccentricity variations, from 0.02 to 0.12. In all cases, the period ratio oscillates around $\sim$ 2.06, slightly above the exact 2:1 period ratio. The angles $\theta_2$ and $\Delta\omega$ circulate between  0$^\circ$ and 360$^\circ$, while $\theta_1$ librates around 0$^\circ$ with a mean semi-amplitude of 73$^\circ$ for all integrated samples. This libration of $\theta_1$ in all samples implies that the massive planets in the TOI-4504 system are involved in a 2:1 MMR, despite the minor offset in period ratio, which, given the strong dynamical interactions, is needed to maintain the resonance.

\autoref{evol_plot} shows an example of a 20-year extent of the orbital evolution simulation started from the best fit (i.e., maximum $-\ln\mathcal{L}$; see \autoref{NS_params}). We show the evolution of the mutual period ratio $P_{\rm rat.}$, the eccentricities $e_{\rm c}$ and $e_{\rm d}$, and the first-order eccentricity-type 2:1 MMR angle $\theta_1$. The evolution of the model based on the best-fit parameters is indicative of the orbital evolution of the posterior samples. The libration of $\theta_1$ is sufficient to conclude that the TOI-4504 Jovian pair is involved in a 2:1 MMR \citep[][]{Lee2004}.

\subsection{Internal composition of TOI-4504\,c}
\begin{figure}
\includegraphics[width=0.72\linewidth,angle=90]{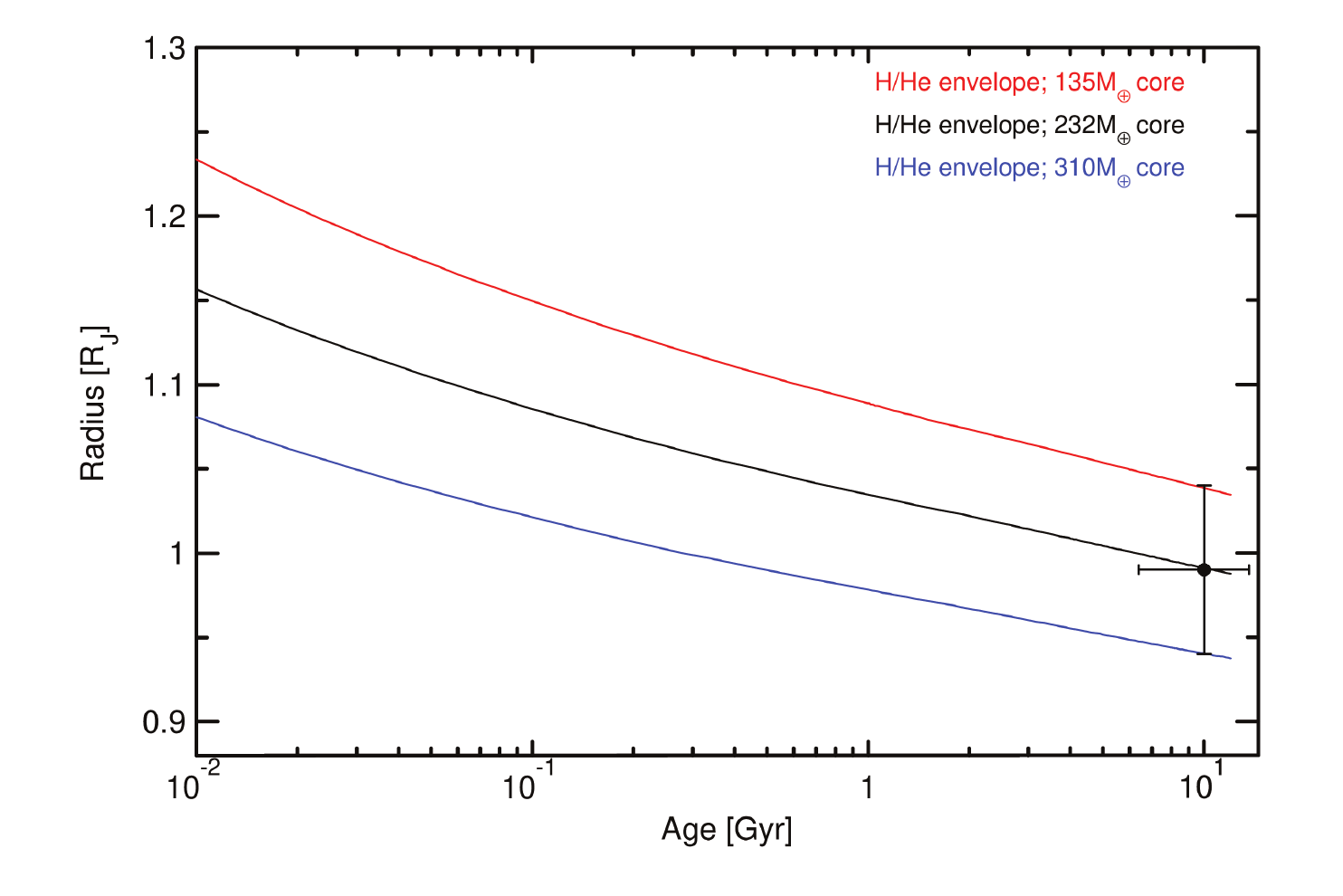}
\caption{Position of TOI-4504\,c in the age-radius diagram (black dot). Three different models with an isodensity core ($\rho$ =  15 [g\,cm$^3$]) with different masses, and surrounded by an H/He envelope are overplotted (solid lines). }
\label{internal_comp} 
\end{figure}

For TOI-4504\,c, we computed planet interior models and their thermal evolution with {\textsc MESA} \citep{mesa1,mesa2}, following the implementation described in \cite{jones2024}. In this case, we modeled the planet with the heavy elements condensed in an inert isodensity core surrounded by a pure gas (H/He) envelope.  
We assumed a 1:1 mixture of rock and ice for the core, with their density obtained from the $\rho-P$ relations 
presented in \cite{hubb1989}. The density of the mixture was computed using the additive volume law. 
We evolved different models with different masses of the core and compared them with the current position of TOI-4504\,c 
in the radius-age diagram. \autoref{internal_comp} shows different models that agree at the 1-$\sigma$ level with the current
radius of the planet. These results correspond to a planet metallicity of Z$_p$  = 0.21$^{+0.07}_{-0.09}$, and a      
corresponding heavy-element enrichment with respect to the host star of Z$_p$/Z$_\star$ = 9.6 $^{+3.1}_{-4.1}$.

\section{Summary and conclusions}
\label{sec5}

In this work, we report the discovery and orbital analysis of a multi-planetary system orbiting the K-dwarf star TOI-4504. We analyzed available photometric data from TESS and conducted follow-up spectroscopic observations with FEROS to constrain the orbital and physical parameters of the detected exoplanets. We confirm two transiting planets, TOI-4504\,b and TOI-4504\,c, and we discover an additional Jovian-mass exoplanet TOI-4504\,d, based on the dynamical perturbations induced on TOI-4504\,c, evidenced by strong TTVs with the largest ever detected semi-amplitude of $\simeq$ 2 days.
Our analysis indicates that the TOI-4504 system consists of a hot sub-Neptune and two warm Jovian planets in a 2:1 MMR. 

The innermost planet, TOI-4504\,b, was only detected in the TESS photometry with a period of 2.42614$^{+0.00014}_{-0.00013}$ days and an estimated radius of $2.69 \pm 0.19$ R$_{\oplus}$. The precision of our RVs is not enough to measure the mass of this potential $\sim$10 M$_{\oplus}$ sub-Neptune \citep{muller:2024}.
TOI-4504\,b is a potentially interesting candidate for further 
characterization of its possible atmospheric composition with JWST \citep[e.g.,][]{madhusudhan:2023,holmberg:2024}. TOI-4504\,b has a predicted transmission spectroscopy metric of $\sim$20 \citep{tsm} and is among the handful of hot sub-Neptunes transiting solar type stars that can be observed with NIRSpec/Prism avoiding saturation, which allows obtaining a transmission spectrum with a wide wavelength coverage (0.6--5.3 $\mu m$ s) from a single transit.


TOI-4504\,c was detected in the TESS data and FEROS RVs and has a period of 82.5438$_{-0.0176}^{+0.0150}$ days valid for epoch BJD = 2458400, a mean orbital period of  82.8540$_{-0.0010}^{+0.0009}$\,days, an estimated radius of $0.9897 \pm 0.0092$\,R$_{\rm Jup}$, and dynamical mass estimate of 3.7672$_{-0.1822}^{+0.1810}$\,M$_{\rm Jup}$. TOI-4504\,c exhibits very large TTVs with a super-period of $\sim$2.9 years and peak-to-peak amplitude of $\sim$4 days. 

Our orbital analysis was focused only on the TTV+RV data for TOI-4504\,c and its non-transiting perturbing planet TOI-4504\,d.
Using a self-consistent TTV+RV model scheme coupled with various optimization and sampling techniques, we were able to pinpoint TOI-4504\,d's period to be 40.5634$_{-0.0368}^{+0.0363}$\,days valid for epoch BJD = 2458400,  a mean orbital period of 40.1716$_{-0.0158}^{+0.0145}$\,days, and a dynamical mass of 1.4166$_{-0.0647}^{+0.0651}$\,M$_{\rm Jup}$.


TESS has already found several strong TTV systems around relatively bright stars, which were effectively followed with ground-based spectroscopic facilities to measure their masses in conjunction with their TTV signals. 
Combining precise RV and TTV data allows for a refined determination of the planetary masses and the system's geometry and dynamical state, aiding in understanding their formation and evolution.
One of these systems is TOI-216 \citep{TOI-216-1, TOI-216-2}. This system hosts a pair of warm gas giants librating at the 2:1 MMR. Another pair of warm giant planets close to 2:1 resonance orbit K-type star TOI-2202 \citep{TOI-2202}. TOI-2525 is another K-type star with two warm giants near the 2:1 MMR \citep{TOI-2525}. TESS observed just a single transit in TIC 279401253, but follow-up RV data are well-fit by a pair of eccentric super-Jupiters in the 2:1 resonance, which would imply large TTVs \citep{Bozhilov2023}.
However, in some respects, the TOI-4504 system shares most similarities with the non-transiting nearby M-dwarf system GJ 876, which was discovered by RVs \citep{GJ876_2,GJ876_1}. GJ 876 hosts a hot super-Earth planet and three other planets trapped in 1:2:4 MMR, a pair of gas-giants (in 2:1 MMR), with the outer planet in the chain being Neptune-mass.

In the context of the GJ\,876 system, the massive resonant pairs in the TOI-4504 and GJ\,876 likely reached their current orbits via slow, convergent type-II migration, combined with planet-planet interactions \citep[see, e.g.,][]{Lee2002, Kley2012}. Similar to systems such as TOI-216, TOI-2202, TOI-2525, and particularly GJ\,876, the TOI-4504 system supports the planet formation and subsequent migration theories, which effectively explain the observed MMR geometries of massive exoplanets.

\begin{figure}[ht!]
  \centering
\includegraphics[width=0.48\textwidth]{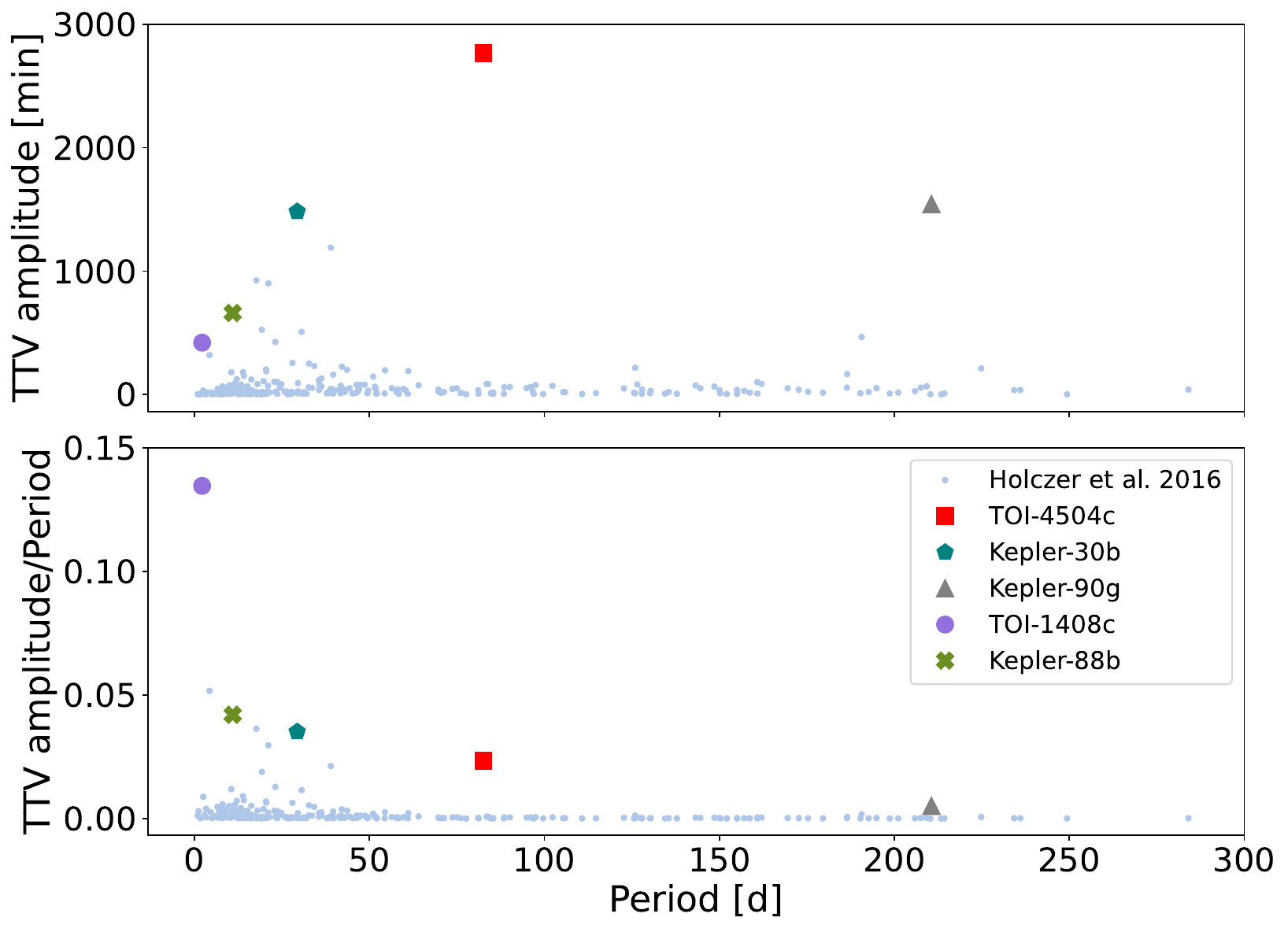}
  \caption{Position of TOI-4504\,c, other planets with significant TTVs and planets from \cite{Holczer2016} in period-TTV amplitude and period-TTV amplitude and period ratio diagram. TTV amplitude is a peak-to-node amplitude of cosinus fit, for Kepler-90\,g it is the maximum observed difference between the time lapsed between consecutive transits.}
  \label{holczer}
\end{figure}

\cite{Holczer2016} made a transit timing catalogue of Kepler Objects of Interest (KOIs) and sorted out several KOIs that showed significant TTVs with long-term variations (see Table 5 in \cite{Holczer2016}. \autoref{holczer} shows TOI-4504\,c in context with other planets with significant TTVs. As can be seen from the top panel of \autoref{holczer}, the peak-to-node amplitude of TTVs of TOI-4504\,c is about twice as big as of Kepler-30\,b, a planet with the largest previously known TTV semi-amplitude \citep{Kepler30b}. However, when calculating the relative amplitude of the TTV with respect to the orbital period of the planet, TOI-4504\,c is not the record holder, although it still belongs to a small group of planets with large relative TTVs (the bottom panel of \autoref{holczer}).

The planet with the largest known TTVs relative to its orbital period is TOI-1408\,c, a sub-Neptune on a very short orbital period \citep{toi-1408}. Our dynamical solution implies that an observer located such that planet d was observed to transit (which is more than 1.6 times as likely as our view of planet c transiting) would observe TTVs more than 50\% larger than those that TESS observed for planet c. Thus, the ratio of d's TTVs to its orbital period would be almost as large as those of hot sub-Neptune TOI-1408 c \citep{toi-1408}.


Additional observations are needed to explain the large RV jitter observed in TOI-4504 and refine the orbital and physical parameters of the system. 
Although our fit to the TTVs is very good, additional observations are needed to further refine the orbital architecture of the TOI-4504\,c and d pair. Transits of TOI-4504\,c Nr. 29 and 31 (\autoref{Tab:TransitsPrediction}) will be observed by TESS in sectors 89 and 95, respectively. Due to significant uncertainties in the predicted transits, ground-based observations will be difficult. Observing campaigns on the nights around the predicted transits would be appreciated. TOI-4504 is within the southern PLATO field and should be monitored continuously for 2 years beginning in 2027 (transits 38 and further).

Since our RV model could not fully resolve the large RV scatter around the fit, the source of the RV jitter remains unclear. The current RV data conclusively confirm the presence of two resonant planets. However, more precise RV measurements are needed to determine the mass of the innermost planet, further constrain their orbital parameters and eccentricities of TOI-4504\,d and TOI-4504\,c, and eventually reveal the presence of additional planets in the system.

Since TOI-4504 is a rather faint target for the 2.2-m telescope with FEROS at La Silla, more precise RVs are urgently needed with facilities like ESPRESSO \citep{Pepe2021} to measure the planetary masses and eccentricities more accurately. We plan to continue our monitoring of TOI-4504 with transit and RV measurements, which will allow us to extend our orbital analysis, capturing the gravitational effects of the planets across the full transit light curve using a photo-dynamical model. These models can measure transit depths and durations, thereby better constraining the dynamic state of the system and shedding more light on the system's origin and migration processes.


\section{Acknowledgments}
We thank the anonymous referee for his/her comments that helped to improve the manuscript. M.V., M.S., and P.K. acknowledge support by Inter-transfer grant no LTT-20015. M.V. also acknowledges travel subsidy from grants ANID-23-05 and EU MSCA EXOWORLD project ID:101086149.
R.B. acknowledges support from FONDECYT Project 1241963 and from ANID -- Millennium  Science  Initiative -- ICN12\_009.
T.T. acknowledges support by the BNSF program "VIHREN-2021" project No. KP-06-DV/5.
M. T. P. acknowledges the support of Fondecyt-ANID fellowship no. 3210253 and ANID ASTRON-0037 project.
A.J.\ acknowledges support from ANID -- Millennium  Science  Initiative -- ICN12\_009 and AIM23-0001 and FONDECYT project 1210718.
K. A. C. acknowledges support from the TESS mission via subaward s3449 from MIT.
This work was funded by the Data Observatory Foundation.
This work was funded by ANID Vinculación Internacional FOVI220091.
This research has made use of the Exoplanet Follow-up Observation Program (ExoFOP; DOI: 10.26134/ExoFOP5) website, which is operated by the California Institute of Technology, under contract with the National Aeronautics and Space Administration under the Exoplanet Exploration Program.
Funding for the TESS mission is provided by NASA's Science Mission Directorate. We acknowledge the use of public TESS data from pipelines at the TESS Science Office and at the TESS Science Processing Operations Center. This paper includes data collected by the TESS mission that are publicly available from the Mikulski Archive for Space Telescopes (MAST). The specific observations analyzed can be accessed via \dataset[https://doi.org/10.17909/g52w-sk09]{https://doi.org/10.17909/g52w-sk09}. STScI is operated by the Association of Universities for Research in Astronomy, Inc., under NASA contract NAS5–26555. Support to MAST for these data is provided by the NASA Office of Space Science via grant NAG5–7584 and by other grants and contracts.
The results reported herein benefitted from collaborations and/or information exchange within NASA's Nexus for Exoplanet System Science (NExSS) research coordination network sponsored by NASA's Science Mission Directorate under Agreement No. 80NSSC21K0593 for the program "Alien Earths".
Resources supporting this work were provided by the NASA High-End Computing (HEC) Program through the NASA Advanced Supercomputing (NAS) Division at Ames Research Center for the production of the SPOC data products.

%

\vspace{5mm}
\facilities{TESS, MPG-2.2m/FEROS}


\software{{\textsc Exo-Striker} \citep{exostriker}, {\textsc juliet} \citep{juliet}, {\textsc ceres} \citep{ceres}, {\textsc zaspe} \citep{zaspe}, {\textsc tesseract} (Rojas, in prep.), {\textsc TESSCut} \citep{tesscut}, {\textsc lightkurve} \citep{lightkurve}, {\textsc dynesty} \citep{dynesty},
{\textsc batman} \citep{batman}, {\textsc celerite} \citep{celerite}}




\appendix

\setcounter{table}{0}
\renewcommand{\thetable}{A\arabic{table}}

\setcounter{figure}{0}
\renewcommand{\thefigure}{A\arabic{figure}}



\begin{table*}[ht]

\centering
\caption{{Priors for the global Nested Sampling run, which aimed to identify the parameter space of TOI-4504\,d consistent with the TTVs of TOI4504\,c and RVs of TOI-4504.}}
\label{globalNS}

\begin{tabular}{lrrrrrrrr}     

\hline\hline  \noalign{\vskip 0.7mm}
Parameter \hspace{0.0 mm}& Planet d & Planet c \\
\hline \noalign{\vskip 0.7mm}

    $K$ [m\,s$^{-1}$]             & $\mathcal{U}$(     30.0,    120.0)& $\mathcal{U}$(    100.0,    250.0)\\
    $P$ [day]                     & $\mathcal{U}$(     20.0,     60.0) \& $\mathcal{U}$( 120.0,     300.0)& $\mathcal{U}$(     81.0,     83.0)\\
    $e\sin(\omega)$               & $\mathcal{U}$(     -0.10,      0.10)& $\mathcal{U}$(     -0.10,      0.10)\\
    $e\cos(\omega)$               & $\mathcal{U}$(     -0.10,      0.10)& $\mathcal{U}$(     -0.10,      0.10)\\
    $\lambda$ [deg]               & $\mathcal{U}$(   -180.0,    360.0)& $\mathcal{U}$(     70.0,    110.0)\\
    $i$ [deg]                     & $\mathcal{N}$(      86.0,    3.0)& $\mathcal{U}$(      89.7,    0.1)\\
    $\Omega$ [deg]                & 0 (fixed) & $\mathcal{N}$(      0.0,    30.0)\\

    RV off.$_{\rm FEROS}$  [m\,s$^{-1}$]& $\mathcal{U}$(   1950.0,   2150.0)\\
    RV jit.$_{\rm FEROS}$  [m\,s$^{-1}$]& $\mathcal{U}$(      0.0,    150.0)\\
    \\
\hline \noalign{\vskip 0.7mm}
    
\end{tabular}



\end{table*}

\section{Upcoming transits of TOI-4504 c}
\label{sec:prediction}
The N-body simulation, together with the best-fit solution of TTVs+RV analysis, allowed us to predict future transit times of TOI-4504\,c more than 10 years in advance \autoref{Tab:TransitsPrediction}. For completeness, we give also past transits since $t_{0}$ in \autoref{Tab:TransitsPrediction}. 

\begin{table*}
\centering
\caption{Predicted mid-transit times for TOI-4504\,c.}
\label{Tab:TransitsPrediction}
\begin{tabular}{cc | cc | cc | cc}
\hline \hline
Nr. & BJD & Nr. & BJD & Nr. & BJD & Nr. & BJD \\ \hline
1  & 2458401.4086$_{-0.0005}^{+0.0004}$ & 21 & 2460059.6184$_{-0.4373}^{+0.3886}$ & 41 & 2461715.7660$_{-1.2393}^{+1.0841}$ & 61 & 2463371.0909$_{-0.9278}^{+0.7731}$ \\
2  & 2458483.2231$_{-0.0153}^{+0.0148}$ & 22 & 2460142.5931$_{-0.3086}^{+0.2823}$ & 42 & 2461799.5893$_{-1.2186}^{+1.0956}$ & 62 & 2463454.1663$_{-1.4180}^{+1.1786}$ \\
3  & 2458565.1073$_{-0.0265}^{+0.0261}$ & 23 & 2460224.9812$_{-0.1876}^{+0.1716}$ & 43 & 2461883.1822$_{-1.0472}^{+0.9626}$ & 63 & 2463537.7434$_{-1.7736}^{+1.5151}$ \\
4  & 2458647.3635$_{-0.0406}^{+0.0390}$ & 24 & 2460306.9802$_{-0.1171}^{+0.1116}$ & 44 & 2461966.2880$_{-0.7521}^{+0.6963}$ & 64 & 2463621.5822$_{-1.8876}^{+1.6493}$ \\
5  & 2458730.2564$_{-0.0845}^{+0.0804}$ & 25 & 2460388.8415$_{-0.1206}^{+0.1139}$ & 45 & 2462048.8016$_{-0.4377}^{+0.4084}$ & 65 & 2463705.3773$_{-1.7615}^{+1.5622}$ \\
6  & 2458813.7926$_{-0.1493}^{+0.1328}$ & 26 & 2460470.8214$_{-0.2082}^{+0.1758}$ & 46 & 2462130.8606$_{-0.2422}^{+0.2417}$ & 66 & 2463788.8495$_{-1.4355}^{+1.3020}$ \\
7  & 2458897.6602$_{-0.1985}^{+0.1770}$ & 27 & 2460553.1805$_{-0.3916}^{+0.3396}$ & 47 & 2462212.7309$_{-0.2153}^{+0.1999}$ & 67 & 2463871.8038$_{-0.9814}^{+0.9202}$ \\
8  & 2458981.5652$_{-0.2339}^{+0.2063}$ & 28 & 2460636.1081$_{-0.6330}^{+0.5527}$ & 48 & 2462294.6947$_{-0.3225}^{+0.2820}$ & 68 & 2463954.2377$_{-0.6102}^{+0.5587}$ \\
9  & 2459065.2622$_{-0.2350}^{+0.2083}$ & 29 & 2460719.5920$_{-0.8437}^{+0.7512}$ & 49 & 2462377.0283$_{-0.5883}^{+0.5114}$ & 69 & 2464036.3360$_{-0.4052}^{+0.3851}$ \\
10 & 2459148.4917$_{-0.1932}^{+0.1722}$ & 30 & 2460803.4249$_{-0.9630}^{+0.8599}$ & 50 & 2462459.9463$_{-0.9861}^{+0.8540}$ & 70 & 2464118.3598$_{-0.4653}^{+0.3946}$ \\
11 & 2459231.1206$_{-0.1300}^{+0.1187}$ & 31 & 2460887.3192$_{-0.9664}^{+0.8721}$ & 51 & 2462543.4456$_{-1.3014}^{+1.1507}$ & 71 & 2464200.5739$_{-0.7364}^{+0.5998}$ \\
12 & 2459313.2552$_{-0.0768}^{+0.0690}$ & 32 & 2460970.9786$_{-0.8496}^{+0.7765}$ & 52 & 2462627.2792$_{-1.4381}^{+1.2986}$ & 72 & 2464283.1879$_{-1.2111}^{+1.0083}$ \\
13 & 2459395.1287$_{-0.0530}^{+0.0510}$ & 33 & 2461054.1342$_{-0.6253}^{+0.5618}$ & 53 & 2462711.1361$_{-1.4059}^{+1.2869}$ & 73 & 2464366.2819$_{-1.7046}^{+1.4592}$ \\
14 & 2459476.9771$_{-0.0608}^{+0.0582}$ & 34 & 2461136.6787$_{-0.3664}^{+0.3407}$ & 54 & 2462794.7524$_{-1.2161}^{+1.1275}$ & 74 & 2464449.7853$_{-2.1040}^{+1.8135}$ \\
15 & 2459559.0306$_{-0.1041}^{+0.0975}$ & 35 & 2461218.7611$_{-0.2035}^{+0.2007}$ & 55 & 2462877.9109$_{-0.8877}^{+0.8377}$ & 75 & 2464533.5273$_{-2.2332}^{+2.0046}$ \\
16 & 2459641.5272$_{-0.1941}^{+0.1779}$ & 36 & 2461300.6731$_{-0.1945}^{+0.1820}$ & 56 & 2462960.5230$_{-0.5423}^{+0.5304}$ & 76 & 2464617.2672$_{-2.1452}^{+1.9669}$ \\
17 & 2459724.6306$_{-0.3385}^{+0.2897}$ & 37 & 2461382.7407$_{-0.3265}^{+0.2812}$ & 57 & 2463042.6845$_{-0.3300}^{+0.3270}$ & 77 & 2464700.7314$_{-1.7739}^{+1.6552}$ \\
18 & 2459808.2733$_{-0.4565}^{+0.3962}$ & 38 & 2461465.2669$_{-0.6066}^{+0.5158}$ & 58 & 2463124.5967$_{-0.2574}^{+0.2358}$ & 78 & 2464783.7092$_{-1.2096}^{+1.1508}$ \\
19 & 2459892.1846$_{-0.5243}^{+0.4561}$ & 39 & 2461548.3851$_{-0.9165}^{+0.7916}$ & 59 & 2463206.4837$_{-0.3098}^{+0.2714}$ & 79 & 2464866.1730$_{-0.7025}^{+0.6756}$ \\
20 & 2459976.0671$_{-0.5178}^{+0.4589}$ & 40 & 2461631.9600$_{-1.1403}^{+0.9770}$ & 60 & 2463288.5747$_{-0.5271}^{+0.4398}$ & 80 & 2464948.2903$_{-0.4299}^{+0.4509}$\\ \hline \hline
\end{tabular}
\tablecomments{Nr gives the number of the transit since t$_0$, and BJD gives the mid-transit times.}
\end{table*}


\begin{table*}
\begin{rotatetable*}
\begin{center}
\caption{Radial velocity and activity indices of TOI-4504 measured with FEROS.}
\label{RV_feros}
\scalebox{0.9}{
\begin{tabular}{ccccccccc}

\hline \hline
BJD             & RV [m/s]           & Bisector     & FWHM    & SNR & $\rm{H}_{\alpha}$      & $\log(\rm{R_{HK}})$  & Na\,II             & He\,I                            \\ \hline 
2458912.730428  & $2023.5 \pm 10.1$ & $-48 \pm 15 $ & 9.9507  & 49     & $0.1554 \pm 0.004  $& $-4.4426 \pm 0.0641 $&$ 0.2966 \pm 0.0073 $&$ 0.5121 \pm 0.0097 $\\
2458917.7301462 & $1774.9 \pm 11.7$ & $-39 \pm 17 $ & 9.9237  & 41     & $0.1346 \pm 0.0043 $& $-5.4889 \pm 0.6482 $&$ 0.3188 \pm 0.0086 $&$ 0.5219 \pm 0.011  $\\
2458925.5538544 & $1992.5 \pm 26.5$ & $75 \pm 34  $ & 10.0707 & 17     & $0.2872 \pm 0.023  $& $-4.1016 \pm 0.0692 $&$ 0.5544 \pm 0.0441 $&$ 0.6636 \pm 0.0436 $\\
2458925.5970903 & $1869.9 \pm 11.3$ & $-66 \pm 16 $ & 9.9409  & 42     & $0.1503 \pm 0.0045 $& $-4.9067 \pm 0.1358 $&$ 0.2471 \pm 0.0076 $&$ 0.536 \pm 0.0106  $\\
2459189.7431016 & $2247.1 \pm 10$   & $-46 \pm 15 $ & 9.8113  & 47     & $0.1494 \pm 0.0042 $& $-5.2786 \pm 0.1929 $&$ 0.2439 \pm 0.0069 $&$ 0.5176 \pm 0.01   $\\
2459192.7143168 & $2255.4 \pm 14.5$ & $-79 \pm 20 $ & 9.7936  & 32     & $0.1557 \pm 0.0073 $& $-4.706 \pm 0.1078  $&$ 0.2373 \pm 0.0124 $&$ 0.4613 \pm 0.0152 $\\
2459207.7670343 & $2210.2 \pm 10.6$ & $-4 \pm 15  $ & 9.8256  & 45     & $0.1463 \pm 0.0043 $& $-5.0808 \pm 0.1269 $&$ 0.1944 \pm 0.0069 $&$ 0.5163 \pm 0.0102 $\\
2459211.7211734 & $2228.6 \pm 10.9$ & $21 \pm 16  $ & 9.8664  & 44     & $0.1596 \pm 0.0047 $& $-5.0505 \pm 0.1226 $&$ 0.2142 \pm 0.0073 $&$ 0.5216 \pm 0.0104 $\\
2459218.7262622 & $2038.1 \pm 10$   & $-47 \pm 15 $ & 9.8061  & 49     & $0.1296 \pm 0.0036 $& $-4.9215 \pm 0.1023 $&$ 0.2176 \pm 0.0061 $&$ 0.5052 \pm 0.0088 $\\
2459260.6382415 & $1877.8 \pm 12$   & $-14 \pm 17 $ & 9.8243  & 39     & $0.1857 \pm 0.0058 $& $-5.2035 \pm 0.2064 $&$ 0.2347 \pm 0.0086 $&$ 0.513 \pm 0.0116  $\\
2459272.776905  & $2081.6 \pm 13.4$ & $4 \pm 19   $ & 9.8138  & 35     & $0.1369 \pm 0.0055 $& $-4.5168 \pm 0.0767 $&$ 0.2836 \pm 0.0105 $&$ 0.5342 \pm 0.0138 $\\
2459274.716173  & $2224.5 \pm 10.6$ & $-1 \pm 15  $ & 9.7323  & 45     & $0.1372 \pm 0.0042 $& $-5.4107 \pm 0.3654 $&$ 0.266 \pm 0.0074  $&$ 0.5071 \pm 0.0097 $\\
2459276.7156498 & $2134.9 \pm 12.9$ & $-46 \pm 18 $ & 10.0024 & 36     & $0.1444 \pm 0.0048 $& $-999 \pm -999      $&$ 0.2532 \pm 0.0087 $&$ 0.5305 \pm 0.0119 $\\
2459282.5358342 & $2267.5 \pm 10.6$ & $11 \pm 15  $ & 9.8555  & 45     & $0.1412 \pm 0.0041 $& $-5.1246 \pm 0.145  $&$ 0.2665 \pm 0.0074 $&$ 0.5166 \pm 0.0103 $\\
2459539.7731663 & $2231.6 \pm 10.9$ & $48 \pm 16  $ & 10.0306 & 44     & $0.1836 \pm 0.0047 $& $-4.5648 \pm 0.0608 $&$ 0.1945 \pm 0.007  $&$ 0.5304 \pm 0.0105 $\\
2459545.7659941 & $2043.8 \pm 12.8$ & $-15 \pm 18 $ & 9.5975  & 36     & $0.238 \pm 0.0062  $& $-6.2329 \pm 3.4563 $&$ 0.2811 \pm 0.01   $&$ 0.5237 \pm 0.0123 $\\
2459590.7117932 & $1786 \pm 13.9$   & $41 \pm 19  $ & 9.807   & 33     & $0.2119 \pm 0.0067 $& $-4.6104 \pm 0.1074 $&$ 0.2212 \pm 0.0112 $&$ 0.5304 \pm 0.0152 $\\
2459644.7122898 & $2181.7 \pm 14.5$ & $-104 \pm 20$ & 9.5813  & 32     & $0.2234 \pm 0.0066 $& $-4.6445 \pm 0.2152 $&$ 0.3935 \pm 0.0129 $&$ 0.5021 \pm 0.0141 $\\
2459649.629842  & $2140.1 \pm 10$   & $-118 \pm 15 $& 9.8357  & 49     & $0.1979 \pm 0.0041 $& $-999 \pm -999      $&$ 0.2381 \pm 0.0065 $&$ 0.4922 \pm 0.0091 $\\
2459652.6796063 & $2016 \pm 12.1$   & $104 \pm 17  $& 9.7958  & 39     & $0.2288 \pm 0.0057 $& $-4.5995 \pm 0.0931 $&$ 0.3102 \pm 0.0094 $&$ 0.5146 \pm 0.0116 $\\
2459657.727332  & $1724.3 \pm 13$   & $-101 \pm 18$ & 10.1905 & 36     & $0.1698 \pm 0.005  $& $-999 \pm -999     $ &$ 0.3019 \pm 0.0097 $&$ 0.5582 \pm 0.0127 $\\
2459661.6841378 & $1910.5 \pm 16.9$ & $-109 \pm 18$& 10.0861 & 38     & $0.24 \pm 0.0062   $& $-999 \pm -999      $& $0.2641 \pm 0.0095 $& $0.538 \pm 0.0127  $\\
2459662.6697007 & $1817.8 \pm 15.2$ & $83 \pm 21  $ & 10.0149 & 30     & $0.2389 \pm 0.0064 $& $-999 \pm -999      $&$ 0.3149 \pm 0.0112 $&$ 0.5291 \pm 0.0141 $\\
2459685.5096644 & $2094.7 \pm 11.7$ & $-106 \pm 17$ & 9.8454  & 41     & $0.1575 \pm 0.0047 $& $-5.2897 \pm 0.36   $&$ 0.3526 \pm 0.0094 $&$ 0.5028 \pm 0.0106 $\\
2459687.5617665 & $2190.1 \pm 10.1$ & $-120 \pm 15$ & 10.0629 & 49     & $0.1979 \pm 0.0048 $& $-4.9255 \pm 0.1053 $&$ 0.3099 \pm 0.0079 $&$ 0.4911 \pm 0.0098 $\\
2459691.5839601 & $2311.2 \pm 14.5$ & $106 \pm 20 $ & 9.8589  & 32     & $0.2592 \pm 0.0073 $& $-4.8472 \pm 0.2618 $&$ 0.4324 \pm 0.0136 $&$ 0.4913 \pm 0.0144 $\\
2459704.623859  & $2092 \pm 15.2$   & $-148 \pm 21$ & 9.914   & 30     & $0.1667 \pm 0.0054 $& $-999 \pm -999      $&$ 0.5008 \pm 0.0121 $&$ 0.5023 \pm 0.0128 $\\
2460031.6812879 & $2174.8 \pm 11.7$ & $0 \pm 17   $ & 9.9468  & 41     & $0.2287 \pm 0.006  $& $-999 \pm -999      $&$ 0.2467 \pm 0.009  $&$ 0.5173 \pm 0.0123 $\\
2460035.6102046 & $2142.3 \pm 9.8$  & $-60 \pm 15 $ & 9.8799  & 50     & $0.1402 \pm 0.0047 $& $-4.7232 \pm 0.0919 $&$ 0.239 \pm 0.006   $&$ 0.5235 \pm 0.0088 $\\
2460047.6196292 & $2185.9 \pm 11.3$ & $-29 \pm 16 $ & 9.8589  & 42     & $0.1388 \pm 0.0075 $& $-6.2879 \pm 4.636  $&$ 0.2799 \pm 0.0081 $&$ 0.5235 \pm 0.0109 $\\
2460049.6297281 & $2235.2 \pm 11.7$ & $-11 \pm 17 $ & 9.9268  & 41     & $0.1787 \pm 0.0048 $& $-4.5127 \pm 0.133  $&$ 0.2593 \pm 0.0081 $&$ 0.5182 \pm 0.0105 $\\
2460053.5979115 & $2273 \pm 10.3$   & $-28 \pm 15 $ & 9.8936  & 47     & $0.1627 \pm 0.0042 $& $-4.6646 \pm 0.0912 $&$ 0.2688 \pm 0.0072 $&$ 0.4967 \pm 0.0094 $\\
2460064.5861888 & $2101.8 \pm 14$   & $-7 \pm 19  $ & 10.0588 & 33     & $0.215 \pm 0.0065  $& $-4.6949 \pm 0.1573 $&$ 0.3868 \pm 0.0117 $&$ 0.5458 \pm 0.0134 $\\
2460067.4999667 & $1900.5 \pm 12.5$ & $12 \pm 18  $ & 10.1785 & 38     & $0.1855 \pm 0.0056 $& $-4.7865 \pm 0.1005 $&$ 0.475 \pm 0.0112  $&$ 0.54 \pm 0.0121  $ \\
2460100.5511551 & $1911 \pm 14.1$   & $-213 \pm 18$ & 10.2172 & 36     & $0.198 \pm 0.0073 $ & $-4.0598 \pm 0.0348 $&$ 0.3902 \pm 0.0125 $&$ 0.4311 \pm 0.0126$ \\ 
2460404.6178393 & $1978.3 \pm 10.4$ & $-60 \pm 15$  & 9.9642 & 47 & $0.1620 \pm 0.0044$ &  $-5.0872 \pm 0.2347$ & $0.3396 \pm 0.0081$ & $0.5046 \pm 0.0100$\\
2460407.5764817 & $1909 \pm 11$     & $68 \pm 16$   & 9.8798 & 44 & $0.1485 \pm 0.0044$ & $-5.1868 \pm 0.2541$ & $0.3106 \pm 0.0082$ & $0.5284 \pm 0.0108$\\
2460411.5489370 & $1878.9 \pm 12.9$ & $90 \pm 18$   & 9.9523 & 36 & $0.1684 \pm 0.0054$ & $-5.0893 \pm 0.2633$ & $0.4273 \pm 0.0109$ & $0.4882 \pm 0.0116$\\
2460433.5833358 & $2270.6 \pm 20.6$ & $-159 \pm 28$ & 9.6584 & 21 & $0.3803 \pm 0.0149$ & $-3.9235 \pm 0.0538$ & $2.5052 \pm 0.0535$ & $0.5286 \pm 0.0315$\\
\hline \hline
\end{tabular}}
\end{center}
\end{rotatetable*}
\end{table*}

\begin{figure*}
    
    \includegraphics[width=17cm]{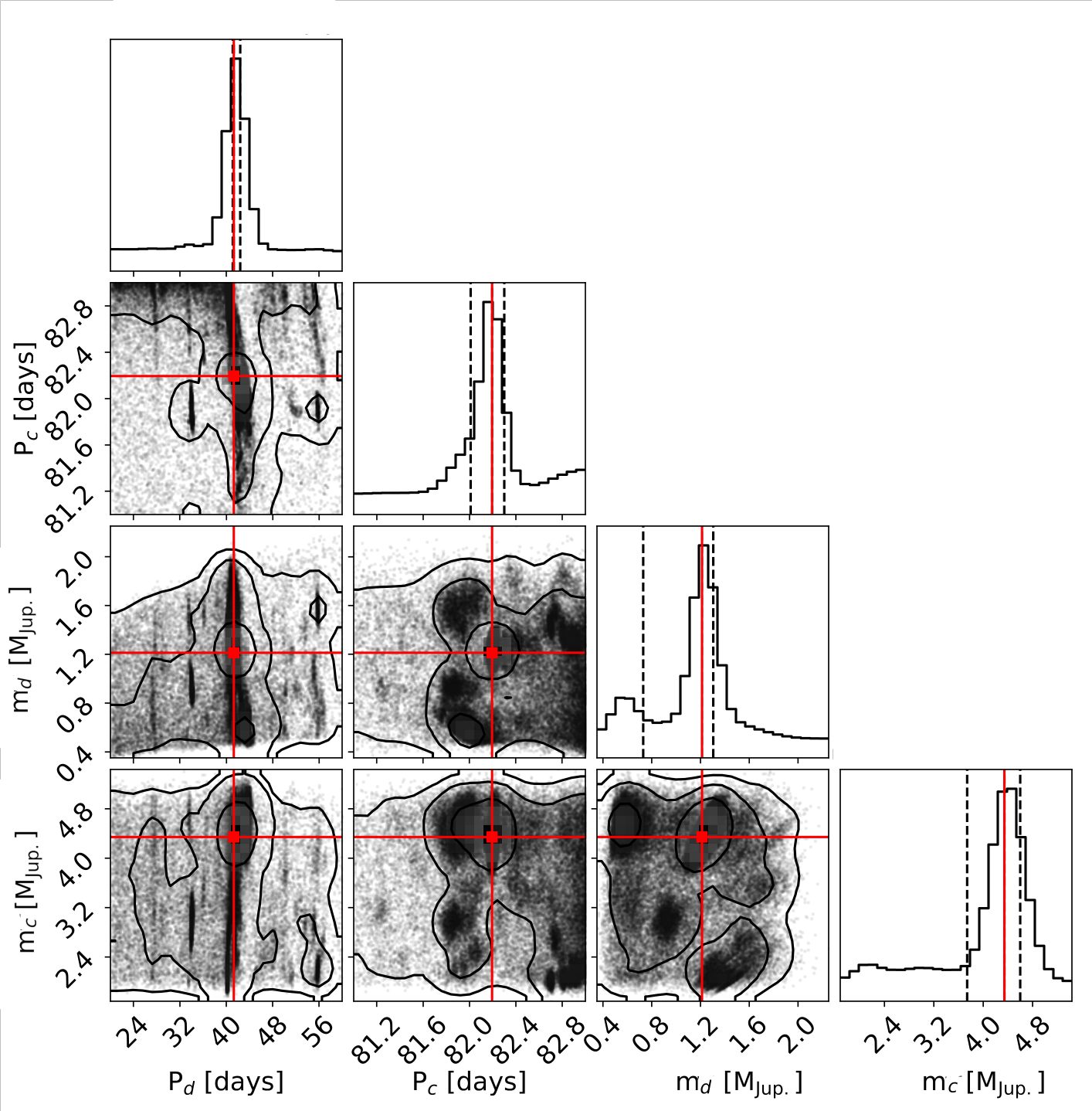}
    \caption{Posterior distribution from a global joint TTV+RV static nested sampling search with very large priors, which aim to map the possible periods for the non-transiting perturber TOI-4504\,d. The posterior is multi-modal implying that different period ratios could produce the observed TTVs, but the 41-day period massive planet leads to significantly better fits pointing towards a 2:1 period ratio commensurability with TOI-4504\,c}
    \label{Nest_samp_ttv}
    
\end{figure*}

\begin{figure*}
    
    \includegraphics[width=17cm]{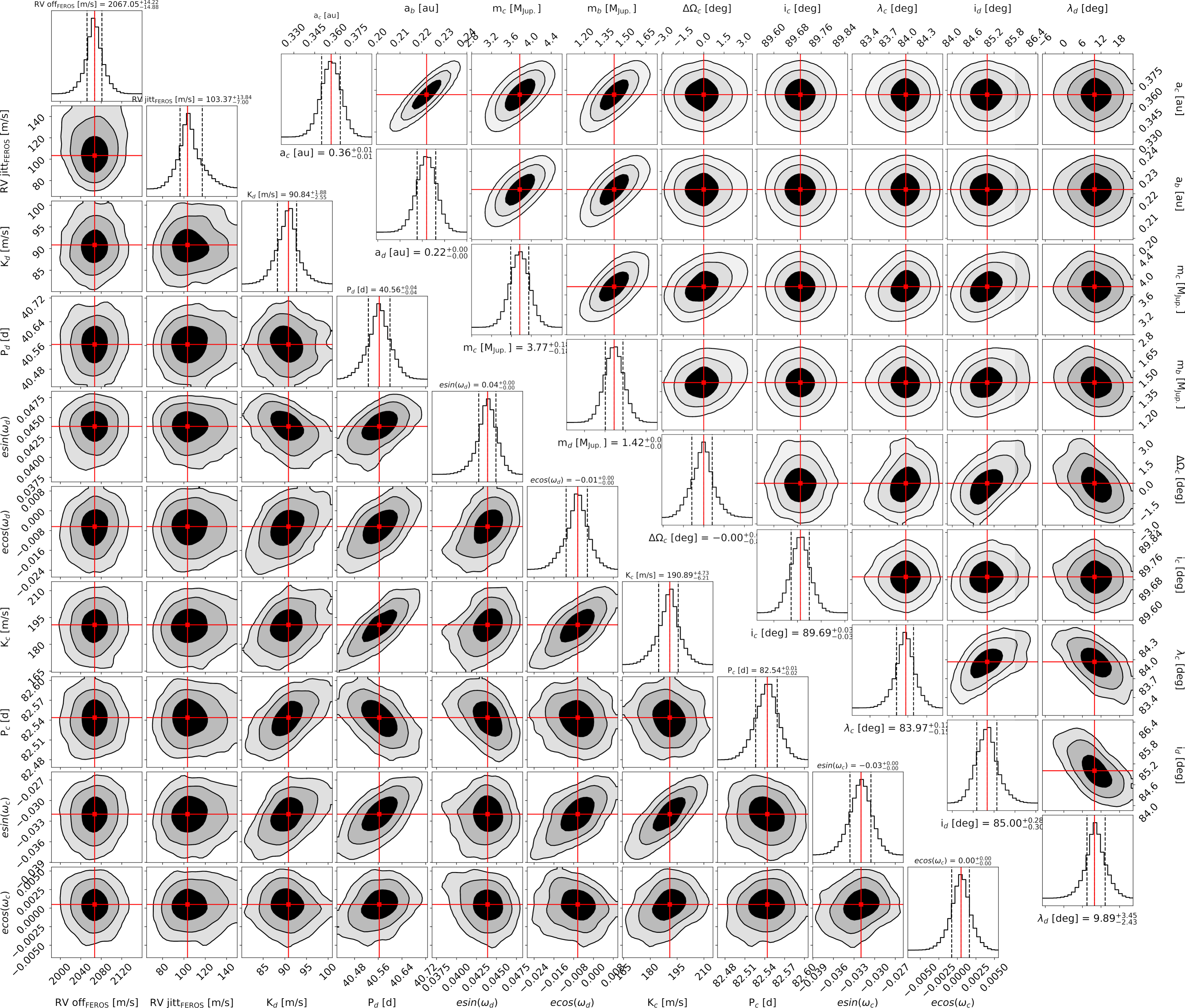}
    \caption{Posterior distribution from a focused joint TTV+RV MCMC sampling, whose results are listed in \autoref{NS_params}. The median values of the fitted and derived posteriors are marked in red. The black contour lines represent the 1,\,2 and 3$\sigma$ intervals of the distribution.}
    
    \label{mcmc_posteriors}
    
\end{figure*} 



\bibliography{sample631}{}

\begin{thebibliography}{}
\expandafter\ifx\csname natexlab\endcsname\relax\def\natexlab#1{#1}\fi
\providecommand{\url}[1]{\href{#1}{#1}}
\providecommand{\dodoi}[1]{doi:~\href{http://doi.org/#1}{\nolinkurl{#1}}}
\providecommand{\doeprint}[1]{\href{http://ascl.net/#1}{\nolinkurl{http://ascl.net/#1}}}
\providecommand{\doarXiv}[1]{\href{https://arxiv.org/abs/#1}{\nolinkurl{https://arxiv.org/abs/#1}}}

\bibitem[{{Agol} {et~al.}(2005){Agol}, {Steffen}, {Sari}, \& {Clarkson}}]{AGOL2005}
{Agol}, E., {Steffen}, J., {Sari}, R., \& {Clarkson}, W. 2005, \mnras, 359, 567, \dodoi{10.1111/j.1365-2966.2005.08922.x}

\bibitem[{{Aller} {et~al.}(2020){Aller}, {Lillo-Box}, {Jones}, {Miranda}, \& {Barcel{\'o} Forteza}}]{tpfplotter}
{Aller}, A., {Lillo-Box}, J., {Jones}, D., {Miranda}, L.~F., \& {Barcel{\'o} Forteza}, S. 2020, \aap, 635, A128, \dodoi{10.1051/0004-6361/201937118}

\bibitem[{{Baluev}(2009)}]{Baluev2009}
{Baluev}, R.~V. 2009, \mnras, 393, 969, \dodoi{10.1111/j.1365-2966.2008.14217.x}

\bibitem[{{Boisse} {et~al.}(2009){Boisse}, {Moutou}, {Vidal-Madjar}, {Bouchy}, {Pont}, {H{\'e}brard}, {Bonfils}, {Croll}, {Delfosse}, {Desort}, {Forveille}, {Lagrange}, {Loeillet}, {Lovis}, {Matthews}, {Mayor}, {Pepe}, {Perrier}, {Queloz}, {Rowe}, {Santos}, {S{\'e}gransan}, \& {Udry}}]{haplha}
{Boisse}, I., {Moutou}, C., {Vidal-Madjar}, A., {et~al.} 2009, \aap, 495, 959, \dodoi{10.1051/0004-6361:200810648}

\bibitem[{{Bozhilov} {et~al.}(2023){Bozhilov}, {Antonova}, {Hobson}, {Brahm}, {Jord{\'a}n}, {Henning}, {Eberhardt}, {Rojas}, {Batygin}, {Torres-Miranda}, {Stassun}, {Millholland}, {Stoeva}, {Minev}, {Espinoza}, {Ricker}, {Latham}, {Dragomir}, {Kunimoto}, {Jenkins}, {Ting}, {Seager}, {Winn}, {Villasenor}, {Bouma}, {Medina}, \& {Trifonov}}]{Bozhilov2023}
{Bozhilov}, V., {Antonova}, D., {Hobson}, M.~J., {et~al.} 2023, \apjl, 946, L36, \dodoi{10.3847/2041-8213/acbd4f}

\bibitem[{{Brahm} {et~al.}(2017{\natexlab{a}}){Brahm}, {Jord{\'a}n}, \& {Espinoza}}]{ceres}
{Brahm}, R., {Jord{\'a}n}, A., \& {Espinoza}, N. 2017{\natexlab{a}}, \pasp, 129, 034002, \dodoi{10.1088/1538-3873/aa5455}

\bibitem[{{Brahm} {et~al.}(2017{\natexlab{b}}){Brahm}, {Jord{\'a}n}, {Hartman}, \& {Bakos}}]{zaspe}
{Brahm}, R., {Jord{\'a}n}, A., {Hartman}, J., \& {Bakos}, G. 2017{\natexlab{b}}, \mnras, 467, 971, \dodoi{10.1093/mnras/stx144}

\bibitem[{Brahm {et~al.}(2019)Brahm, Espinoza, Jordán, Henning, Sarkis, Jones, Díaz, Jenkins, Vanzi, Zapata, Petrovich, Kossakowski, Rabus, Rojas, \& Torres}]{wine1}
Brahm, R., Espinoza, N., Jordán, A., {et~al.} 2019, The Astronomical Journal, 158, 45, \dodoi{10.3847/1538-3881/ab279a}

\bibitem[{{Brasseur} {et~al.}(2019){Brasseur}, {Phillip}, {Fleming}, {Mullally}, \& {White}}]{tesscut}
{Brasseur}, C.~E., {Phillip}, C., {Fleming}, S.~W., {Mullally}, S.~E., \& {White}, R.~L. 2019, {Astrocut: Tools for creating cutouts of TESS images}, Astrophysics Source Code Library, record ascl:1905.007

\bibitem[{{Cabrera} {et~al.}(2014){Cabrera}, {Csizmadia}, {Lehmann}, {Dvorak}, {Gandolfi}, {Rauer}, {Erikson}, {Dreyer}, {Eigm{\"u}ller}, \& {Hatzes}}]{Kepler90}
{Cabrera}, J., {Csizmadia}, S., {Lehmann}, H., {et~al.} 2014, \apj, 781, 18, \dodoi{10.1088/0004-637X/781/1/18}

\bibitem[{{Cardelli} {et~al.}(1989){Cardelli}, {Clayton}, \& {Mathis}}]{cardelli:1989}
{Cardelli}, J.~A., {Clayton}, G.~C., \& {Mathis}, J.~S. 1989, \apj, 345, 245, \dodoi{10.1086/167900}

\bibitem[{Dawson {et~al.}(2019)Dawson, Huang, Lissauer, Collins, Sha, Armstrong, Conti, Collins, Evans, Gan, Horne, Ireland, Murgas, Myers, Relles, Sefako, Shporer, Stockdale, Žerjal, Zhou, Ricker, Vanderspek, Latham, Seager, Winn, Jenkins, Bouma, Caldwell, Daylan, Doty, Dynes, Esquerdo, Rose, Smith, \& Yu}]{TOI-216-1}
Dawson, R.~I., Huang, C.~X., Lissauer, J.~J., {et~al.} 2019, The Astronomical Journal, 158, 65, \dodoi{10.3847/1538-3881/ab24ba}

\bibitem[{Dawson {et~al.}(2021)Dawson, Huang, Brahm, Collins, Hobson, Jordán, Dong, Korth, Trifonov, Abe, Agabi, Bruni, Butler, Barbieri, Collins, Conti, Crane, Crouzet, Dransfield, Evans, Espinoza, Gan, Guillot, Henning, Lissauer, Jensen, Sainte, Mékarnia, Myers, Nandakumar, Relles, Sarkis, Torres, Shectman, Schmider, Shporer, Stockdale, Teske, Triaud, Wang, Ziegler, Ricker, Vanderspek, Latham, Seager, Winn, Jenkins, Bouma, Burt, Charbonneau, Levine, McDermott, McLean, Rose, Vanderburg, \& Wohler}]{TOI-216-2}
Dawson, R.~I., Huang, C.~X., Brahm, R., {et~al.} 2021, The Astronomical Journal, 161, 161, \dodoi{10.3847/1538-3881/abd8d0}

\bibitem[{{Deck} {et~al.}(2014){Deck}, {Agol}, {Holman}, \& {Nesvorn{\'y}}}]{Deck2014}
{Deck}, K.~M., {Agol}, E., {Holman}, M.~J., \& {Nesvorn{\'y}}, D. 2014, \apj, 787, 132, \dodoi{10.1088/0004-637X/787/2/132}

\bibitem[{{Dobrovolskis} \& {Borucki}(1996)}]{TTV1}
{Dobrovolskis}, A.~R., \& {Borucki}, W.~J. 1996, in Bulletin of the American Astronomical Society, Vol.~28, 1112

\bibitem[{{Duncan} {et~al.}(1991){Duncan}, {Vaughan}, {Wilson}, {Preston}, {Frazer}, {Lanning}, {Misch}, {Mueller}, {Soyumer}, {Woodard}, {Baliunas}, {Noyes}, {Hartmann}, {Porter}, {Zwaan}, {Middelkoop}, {Rutten}, \& {Mihalas}}]{log2}
{Duncan}, D.~K., {Vaughan}, A.~H., {Wilson}, O.~C., {et~al.} 1991, \apjs, 76, 383, \dodoi{10.1086/191572}

\bibitem[{{Espinoza} {et~al.}(2018){Espinoza}, {Kossakowski}, \& {Brahm}}]{juliet}
{Espinoza}, N., {Kossakowski}, D., \& {Brahm}, R. 2018, arXiv e-prints, arXiv:1812.08549.
\newblock \doarXiv{1812.08549}

\bibitem[{{Foreman-Mackey} {et~al.}(2017){Foreman-Mackey}, {Agol}, {Angus}, \& {Ambikasaran}}]{celerite}
{Foreman-Mackey}, D., {Agol}, E., {Angus}, R., \& {Ambikasaran}, S. 2017, AJ, 154, 220, \dodoi{10.3847/1538-3881/aa9332}

\bibitem[{{Foreman-Mackey} {et~al.}(2013){Foreman-Mackey}, {Hogg}, {Lang}, \& {Goodman}}]{emcee}
{Foreman-Mackey}, D., {Hogg}, D.~W., {Lang}, D., \& {Goodman}, J. 2013, PASP, 125, 306, \dodoi{10.1086/670067}

\bibitem[{{Gaia Collaboration} {et~al.}(2016){Gaia Collaboration}, {Prusti}, {de Bruijne}, {Brown}, {Vallenari}, {Babusiaux}, {Bailer-Jones}, {Bastian}, {Biermann}, {Evans}, {Eyer}, {Jansen}, {Jordi}, {Klioner}, {Lammers}, {Lindegren}, {Luri}, {Mignard}, {Milligan}, {Panem}, {Poinsignon}, {Pourbaix}, {Randich}, {Sarri}, {Sartoretti}, {Siddiqui}, {Soubiran}, {Valette}, {van Leeuwen}, {Walton}, {Aerts}, {Arenou}, {Cropper}, {Drimmel}, {H{\o}g}, {Katz}, {Lattanzi}, {O'Mullane}, {Grebel}, {Holland}, {Huc}, {Passot}, {Bramante}, {Cacciari}, {Casta{\~n}eda}, {Chaoul}, {Cheek}, {De Angeli}, {Fabricius}, {Guerra}, {Hern{\'a}ndez}, {Jean-Antoine-Piccolo}, {Masana}, {Messineo}, {Mowlavi}, {Nienartowicz}, {Ord{\'o}{\~n}ez-Blanco}, {Panuzzo}, {Portell}, {Richards}, {Riello}, {Seabroke}, {Tanga}, {Th{\'e}venin}, {Torra}, {Els}, {Gracia-Abril}, {Comoretto}, {Garcia-Reinaldos}, {Lock}, {Mercier}, {Altmann}, {Andrae}, {Astraatmadja}, {Bellas-Velidis}, {Benson}, {Berthier}, {Blomme}, {Busso}, {Carry}, {Cellino}, {Clementini},
  {Cowell}, {Creevey}, {Cuypers}, {Davidson}, {De Ridder}, {de Torres}, {Delchambre}, {Dell'Oro}, {Ducourant}, {Fr{\'e}mat}, {Garc{\'\i}a-Torres}, {Gosset}, {Halbwachs}, {Hambly}, {Harrison}, {Hauser}, {Hestroffer}, {Hodgkin}, {Huckle}, {Hutton}, {Jasniewicz}, {Jordan}, {Kontizas}, {Korn}, {Lanzafame}, {Manteiga}, {Moitinho}, {Muinonen}, {Osinde}, {Pancino}, {Pauwels}, {Petit}, {Recio-Blanco}, {Robin}, {Sarro}, {Siopis}, {Smith}, {Smith}, {Sozzetti}, {Thuillot}, {van Reeven}, {Viala}, {Abbas}, {Abreu Aramburu}, {Accart}, {Aguado}, {Allan}, {Allasia}, {Altavilla}, {{\'A}lvarez}, {Alves}, {Anderson}, {Andrei}, {Anglada Varela}, {Antiche}, {Antoja}, {Ant{\'o}n}, {Arcay}, {Atzei}, {Ayache}, {Bach}, {Baker}, {Balaguer-N{\'u}{\~n}ez}, {Barache}, {Barata}, {Barbier}, {Barblan}, {Baroni}, {Barrado y Navascu{\'e}s}, {Barros}, {Barstow}, {Becciani}, {Bellazzini}, {Bellei}, {Bello Garc{\'\i}a}, {Belokurov}, {Bendjoya}, {Berihuete}, {Bianchi}, {Bienaym{\'e}}, {Billebaud}, {Blagorodnova}, {Blanco-Cuaresma}, {Boch},
  {Bombrun}, {Borrachero}, {Bouquillon}, {Bourda}, {Bouy}, {Bragaglia}, {Breddels}, {Brouillet}, {Br{\"u}semeister}, {Bucciarelli}, {Budnik}, {Burgess}, {Burgon}, {Burlacu}, {Busonero}, {Buzzi}, {Caffau}, {Cambras}, {Campbell}, {Cancelliere}, {Cantat-Gaudin}, {Carlucci}, {Carrasco}, {Castellani}, {Charlot}, {Charnas}, {Charvet}, {Chassat}, {Chiavassa}, {Clotet}, {Cocozza}, {Collins}, {Collins}, {Costigan}, {Crifo}, {Cross}, {Crosta}, {Crowley}, {Dafonte}, {Damerdji}, {Dapergolas}, {David}, {David}, {De Cat}, {de Felice}, {de Laverny}, {De Luise}, {De March}, {de Martino}, {de Souza}, {Debosscher}, {del Pozo}, {Delbo}, {Delgado}, {Delgado}, {di Marco}, {Di Matteo}, {Diakite}, {Distefano}, {Dolding}, {Dos Anjos}, {Drazinos}, {Dur{\'a}n}, {Dzigan}, {Ecale}, {Edvardsson}, {Enke}, {Erdmann}, {Escolar}, {Espina}, {Evans}, {Eynard Bontemps}, {Fabre}, {Fabrizio}, {Faigler}, {Falc{\~a}o}, {Farr{\`a}s Casas}, {Faye}, {Federici}, {Fedorets}, {Fern{\'a}ndez-Hern{\'a}ndez}, {Fernique}, {Fienga}, {Figueras}, {Filippi},
  {Findeisen}, {Fonti}, {Fouesneau}, {Fraile}, {Fraser}, {Fuchs}, {Furnell}, {Gai}, {Galleti}, {Galluccio}, {Garabato}, {Garc{\'\i}a-Sedano}, {Gar{\'e}}, {Garofalo}, {Garralda}, {Gavras}, {Gerssen}, {Geyer}, {Gilmore}, {Girona}, {Giuffrida}, {Gomes}, {Gonz{\'a}lez-Marcos}, {Gonz{\'a}lez-N{\'u}{\~n}ez}, {Gonz{\'a}lez-Vidal}, {Granvik}, {Guerrier}, {Guillout}, {Guiraud}, {G{\'u}rpide}, {Guti{\'e}rrez-S{\'a}nchez}, {Guy}, {Haigron}, {Hatzidimitriou}, {Haywood}, {Heiter}, {Helmi}, {Hobbs}, {Hofmann}, {Holl}, {Holland}, {Hunt}, {Hypki}, {Icardi}, {Irwin}, {Jevardat de Fombelle}, {Jofr{\'e}}, {Jonker}, {Jorissen}, {Julbe}, {Karampelas}, {Kochoska}, {Kohley}, {Kolenberg}, {Kontizas}, {Koposov}, {Kordopatis}, {Koubsky}, {Kowalczyk}, {Krone-Martins}, {Kudryashova}, {Kull}, {Bachchan}, {Lacoste-Seris}, {Lanza}, {Lavigne}, {Le Poncin-Lafitte}, {Lebreton}, {Lebzelter}, {Leccia}, {Leclerc}, {Lecoeur-Taibi}, {Lemaitre}, {Lenhardt}, {Leroux}, {Liao}, {Licata}, {Lindstr{\o}m}, {Lister}, {Livanou}, {Lobel}, {L{\"o}ffler},
  {L{\'o}pez}, {Lopez-Lozano}, {Lorenz}, {Loureiro}, {MacDonald}, {Magalh{\~a}es Fernandes}, {Managau}, {Mann}, {Mantelet}, {Marchal}, {Marchant}, {Marconi}, {Marie}, {Marinoni}, {Marrese}, {Marschalk{\'o}}, {Marshall}, {Mart{\'\i}n-Fleitas}, {Martino}, {Mary}, {Matijevi{\v{c}}}, {Mazeh}, {McMillan}, {Messina}, {Mestre}, {Michalik}, {Millar}, {Miranda}, {Molina}, {Molinaro}, {Molinaro}, {Moln{\'a}r}, {Moniez}, {Montegriffo}, {Monteiro}, {Mor}, {Mora}, {Morbidelli}, {Morel}, {Morgenthaler}, {Morley}, {Morris}, {Mulone}, {Muraveva}, {Musella}, {Narbonne}, {Nelemans}, {Nicastro}, {Noval}, {Ord{\'e}novic}, {Ordieres-Mer{\'e}}, {Osborne}, {Pagani}, {Pagano}, {Pailler}, {Palacin}, {Palaversa}, {Parsons}, {Paulsen}, {Pecoraro}, {Pedrosa}, {Pentik{\"a}inen}, {Pereira}, {Pichon}, {Piersimoni}, {Pineau}, {Plachy}, {Plum}, {Poujoulet}, {Pr{\v{s}}a}, {Pulone}, {Ragaini}, {Rago}, {Rambaux}, {Ramos-Lerate}, {Ranalli}, {Rauw}, {Read}, {Regibo}, {Renk}, {Reyl{\'e}}, {Ribeiro}, {Rimoldini}, {Ripepi}, {Riva}, {Rixon},
  {Roelens}, {Romero-G{\'o}mez}, {Rowell}, {Royer}, {Rudolph}, {Ruiz-Dern}, {Sadowski}, {Sagrist{\`a} Sell{\'e}s}, {Sahlmann}, {Salgado}, {Salguero}, {Sarasso}, {Savietto}, {Schnorhk}, {Schultheis}, {Sciacca}, {Segol}, {Segovia}, {Segransan}, {Serpell}, {Shih}, {Smareglia}, {Smart}, {Smith}, {Solano}, {Solitro}, {Sordo}, {Soria Nieto}, {Souchay}, {Spagna}, {Spoto}, {Stampa}, {Steele}, {Steidelm{\"u}ller}, {Stephenson}, {Stoev}, {Suess}, {S{\"u}veges}, {Surdej}, {Szabados}, {Szegedi-Elek}, {Tapiador}, {Taris}, {Tauran}, {Taylor}, {Teixeira}, {Terrett}, {Tingley}, {Trager}, {Turon}, {Ulla}, {Utrilla}, {Valentini}, {van Elteren}, {Van Hemelryck}, {van Leeuwen}, {Varadi}, {Vecchiato}, {Veljanoski}, {Via}, {Vicente}, {Vogt}, {Voss}, {Votruba}, {Voutsinas}, {Walmsley}, {Weiler}, {Weingrill}, {Werner}, {Wevers}, {Whitehead}, {Wyrzykowski}, {Yoldas}, {{\v{Z}}erjal}, {Zucker}, {Zurbach}, {Zwitter}, {Alecu}, {Allen}, {Allende Prieto}, {Amorim}, {Anglada-Escud{\'e}}, {Arsenijevic}, {Azaz}, {Balm}, {Beck}, {Bernstein},
  {Bigot}, {Bijaoui}, {Blasco}, {Bonfigli}, {Bono}, {Boudreault}, {Bressan}, {Brown}, {Brunet}, {Bunclark}, {Buonanno}, {Butkevich}, {Carret}, {Carrion}, {Chemin}, {Ch{\'e}reau}, {Corcione}, {Darmigny}, {de Boer}, {de Teodoro}, {de Zeeuw}, {Delle Luche}, {Domingues}, {Dubath}, {Fodor}, {Fr{\'e}zouls}, {Fries}, {Fustes}, {Fyfe}, {Gallardo}, {Gallegos}, {Gardiol}, {Gebran}, {Gomboc}, {G{\'o}mez}, {Grux}, {Gueguen}, {Heyrovsky}, {Hoar}, {Iannicola}, {Isasi Parache}, {Janotto}, {Joliet}, {Jonckheere}, {Keil}, {Kim}, {Klagyivik}, {Klar}, {Knude}, {Kochukhov}, {Kolka}, {Kos}, {Kutka}, {Lainey}, {LeBouquin}, {Liu}, {Loreggia}, {Makarov}, {Marseille}, {Martayan}, {Martinez-Rubi}, {Massart}, {Meynadier}, {Mignot}, {Munari}, {Nguyen}, {Nordlander}, {Ocvirk}, {O'Flaherty}, {Olias Sanz}, {Ortiz}, {Osorio}, {Oszkiewicz}, {Ouzounis}, {Palmer}, {Park}, {Pasquato}, {Peltzer}, {Peralta}, {P{\'e}turaud}, {Pieniluoma}, {Pigozzi}, {Poels}, {Prat}, {Prod'homme}, {Raison}, {Rebordao}, {Risquez}, {Rocca-Volmerange}, {Rosen},
  {Ruiz-Fuertes}, {Russo}, {Sembay}, {Serraller Vizcaino}, {Short}, {Siebert}, {Silva}, {Sinachopoulos}, {Slezak}, {Soffel}, {Sosnowska}, {Strai{\v{z}}ys}, {ter Linden}, {Terrell}, {Theil}, {Tiede}, {Troisi}, {Tsalmantza}, {Tur}, {Vaccari}, {Vachier}, {Valles}, {Van Hamme}, {Veltz}, {Virtanen}, {Wallut}, {Wichmann}, {Wilkinson}, {Ziaeepour}, \& {Zschocke}}]{gaia_a}
{Gaia Collaboration}, {Prusti}, T., {de Bruijne}, J.~H.~J., {et~al.} 2016, \aap, 595, A1, \dodoi{10.1051/0004-6361/201629272}

\bibitem[{{Gaia Collaboration} {et~al.}(2023){Gaia Collaboration}, {Vallenari}, {Brown}, {Prusti}, {de Bruijne}, {Arenou}, {Babusiaux}, {Biermann}, {Creevey}, {Ducourant}, {Evans}, {Eyer}, {Guerra}, {Hutton}, {Jordi}, {Klioner}, {Lammers}, {Lindegren}, {Luri}, {Mignard}, {Panem}, {Pourbaix}, {Randich}, {Sartoretti}, {Soubiran}, {Tanga}, {Walton}, {Bailer-Jones}, {Bastian}, {Drimmel}, {Jansen}, {Katz}, {Lattanzi}, {van Leeuwen}, {Bakker}, {Cacciari}, {Casta{\~n}eda}, {De Angeli}, {Fabricius}, {Fouesneau}, {Fr{\'e}mat}, {Galluccio}, {Guerrier}, {Heiter}, {Masana}, {Messineo}, {Mowlavi}, {Nicolas}, {Nienartowicz}, {Pailler}, {Panuzzo}, {Riclet}, {Roux}, {Seabroke}, {Sordo}, {Th{\'e}venin}, {Gracia-Abril}, {Portell}, {Teyssier}, {Altmann}, {Andrae}, {Audard}, {Bellas-Velidis}, {Benson}, {Berthier}, {Blomme}, {Burgess}, {Busonero}, {Busso}, {C{\'a}novas}, {Carry}, {Cellino}, {Cheek}, {Clementini}, {Damerdji}, {Davidson}, {de Teodoro}, {Nu{\~n}ez Campos}, {Delchambre}, {Dell'Oro}, {Esquej},
  {Fern{\'a}ndez-Hern{\'a}ndez}, {Fraile}, {Garabato}, {Garc{\'\i}a-Lario}, {Gosset}, {Haigron}, {Halbwachs}, {Hambly}, {Harrison}, {Hern{\'a}ndez}, {Hestroffer}, {Hodgkin}, {Holl}, {Jan{\ss}en}, {Jevardat de Fombelle}, {Jordan}, {Krone-Martins}, {Lanzafame}, {L{\"o}ffler}, {Marchal}, {Marrese}, {Moitinho}, {Muinonen}, {Osborne}, {Pancino}, {Pauwels}, {Recio-Blanco}, {Reyl{\'e}}, {Riello}, {Rimoldini}, {Roegiers}, {Rybizki}, {Sarro}, {Siopis}, {Smith}, {Sozzetti}, {Utrilla}, {van Leeuwen}, {Abbas}, {{\'A}brah{\'a}m}, {Abreu Aramburu}, {Aerts}, {Aguado}, {Ajaj}, {Aldea-Montero}, {Altavilla}, {{\'A}lvarez}, {Alves}, {Anders}, {Anderson}, {Anglada Varela}, {Antoja}, {Baines}, {Baker}, {Balaguer-N{\'u}{\~n}ez}, {Balbinot}, {Balog}, {Barache}, {Barbato}, {Barros}, {Barstow}, {Bartolom{\'e}}, {Bassilana}, {Bauchet}, {Becciani}, {Bellazzini}, {Berihuete}, {Bernet}, {Bertone}, {Bianchi}, {Binnenfeld}, {Blanco-Cuaresma}, {Blazere}, {Boch}, {Bombrun}, {Bossini}, {Bouquillon}, {Bragaglia}, {Bramante}, {Breedt},
  {Bressan}, {Brouillet}, {Brugaletta}, {Bucciarelli}, {Burlacu}, {Butkevich}, {Buzzi}, {Caffau}, {Cancelliere}, {Cantat-Gaudin}, {Carballo}, {Carlucci}, {Carnerero}, {Carrasco}, {Casamiquela}, {Castellani}, {Castro-Ginard}, {Chaoul}, {Charlot}, {Chemin}, {Chiaramida}, {Chiavassa}, {Chornay}, {Comoretto}, {Contursi}, {Cooper}, {Cornez}, {Cowell}, {Crifo}, {Cropper}, {Crosta}, {Crowley}, {Dafonte}, {Dapergolas}, {David}, {David}, {de Laverny}, {De Luise}, {De March}, {De Ridder}, {de Souza}, {de Torres}, {del Peloso}, {del Pozo}, {Delbo}, {Delgado}, {Delisle}, {Demouchy}, {Dharmawardena}, {Di Matteo}, {Diakite}, {Diener}, {Distefano}, {Dolding}, {Edvardsson}, {Enke}, {Fabre}, {Fabrizio}, {Faigler}, {Fedorets}, {Fernique}, {Fienga}, {Figueras}, {Fournier}, {Fouron}, {Fragkoudi}, {Gai}, {Garcia-Gutierrez}, {Garcia-Reinaldos}, {Garc{\'\i}a-Torres}, {Garofalo}, {Gavel}, {Gavras}, {Gerlach}, {Geyer}, {Giacobbe}, {Gilmore}, {Girona}, {Giuffrida}, {Gomel}, {Gomez}, {Gonz{\'a}lez-N{\'u}{\~n}ez},
  {Gonz{\'a}lez-Santamar{\'\i}a}, {Gonz{\'a}lez-Vidal}, {Granvik}, {Guillout}, {Guiraud}, {Guti{\'e}rrez-S{\'a}nchez}, {Guy}, {Hatzidimitriou}, {Hauser}, {Haywood}, {Helmer}, {Helmi}, {Sarmiento}, {Hidalgo}, {Hilger}, {H{\l}adczuk}, {Hobbs}, {Holland}, {Huckle}, {Jardine}, {Jasniewicz}, {Jean-Antoine Piccolo}, {Jim{\'e}nez-Arranz}, {Jorissen}, {Juaristi Campillo}, {Julbe}, {Karbevska}, {Kervella}, {Khanna}, {Kontizas}, {Kordopatis}, {Korn}, {K{\'o}sp{\'a}l}, {Kostrzewa-Rutkowska}, {Kruszy{\'n}ska}, {Kun}, {Laizeau}, {Lambert}, {Lanza}, {Lasne}, {Le Campion}, {Lebreton}, {Lebzelter}, {Leccia}, {Leclerc}, {Lecoeur-Taibi}, {Liao}, {Licata}, {Lindstr{\o}m}, {Lister}, {Livanou}, {Lobel}, {Lorca}, {Loup}, {Madrero Pardo}, {Magdaleno Romeo}, {Managau}, {Mann}, {Manteiga}, {Marchant}, {Marconi}, {Marcos}, {Marcos Santos}, {Mar{\'\i}n Pina}, {Marinoni}, {Marocco}, {Marshall}, {Martin Polo}, {Mart{\'\i}n-Fleitas}, {Marton}, {Mary}, {Masip}, {Massari}, {Mastrobuono-Battisti}, {Mazeh}, {McMillan}, {Messina}, {Michalik},
  {Millar}, {Mints}, {Molina}, {Molinaro}, {Moln{\'a}r}, {Monari}, {Mongui{\'o}}, {Montegriffo}, {Montero}, {Mor}, {Mora}, {Morbidelli}, {Morel}, {Morris}, {Muraveva}, {Murphy}, {Musella}, {Nagy}, {Noval}, {Oca{\~n}a}, {Ogden}, {Ordenovic}, {Osinde}, {Pagani}, {Pagano}, {Palaversa}, {Palicio}, {Pallas-Quintela}, {Panahi}, {Payne-Wardenaar}, {Pe{\~n}alosa Esteller}, {Penttil{\"a}}, {Pichon}, {Piersimoni}, {Pineau}, {Plachy}, {Plum}, {Poggio}, {Pr{\v{s}}a}, {Pulone}, {Racero}, {Ragaini}, {Rainer}, {Raiteri}, {Rambaux}, {Ramos}, {Ramos-Lerate}, {Re Fiorentin}, {Regibo}, {Richards}, {Rios Diaz}, {Ripepi}, {Riva}, {Rix}, {Rixon}, {Robichon}, {Robin}, {Robin}, {Roelens}, {Rogues}, {Rohrbasser}, {Romero-G{\'o}mez}, {Rowell}, {Royer}, {Ruz Mieres}, {Rybicki}, {Sadowski}, {S{\'a}ez N{\'u}{\~n}ez}, {Sagrist{\`a} Sell{\'e}s}, {Sahlmann}, {Salguero}, {Samaras}, {Sanchez Gimenez}, {Sanna}, {Santove{\~n}a}, {Sarasso}, {Schultheis}, {Sciacca}, {Segol}, {Segovia}, {S{\'e}gransan}, {Semeux}, {Shahaf}, {Siddiqui}, {Siebert},
  {Siltala}, {Silvelo}, {Slezak}, {Slezak}, {Smart}, {Snaith}, {Solano}, {Solitro}, {Souami}, {Souchay}, {Spagna}, {Spina}, {Spoto}, {Steele}, {Steidelm{\"u}ller}, {Stephenson}, {S{\"u}veges}, {Surdej}, {Szabados}, {Szegedi-Elek}, {Taris}, {Taylor}, {Teixeira}, {Tolomei}, {Tonello}, {Torra}, {Torra}, {Torralba Elipe}, {Trabucchi}, {Tsounis}, {Turon}, {Ulla}, {Unger}, {Vaillant}, {van Dillen}, {van Reeven}, {Vanel}, {Vecchiato}, {Viala}, {Vicente}, {Voutsinas}, {Weiler}, {Wevers}, {Wyrzykowski}, {Yoldas}, {Yvard}, {Zhao}, {Zorec}, {Zucker}, \& {Zwitter}}]{gaia_b}
{Gaia Collaboration}, {Vallenari}, A., {Brown}, A.~G.~A., {et~al.} 2023, \aap, 674, A1, \dodoi{10.1051/0004-6361/202243940}

\bibitem[{{Gladman}(1993)}]{Gladman1993}
{Gladman}, B. 1993, \icarus, 106, 247, \dodoi{10.1006/icar.1993.1169}

\bibitem[{{Gomes da Silva} {et~al.}(2011){Gomes da Silva}, {Santos}, {Bonfils}, {Delfosse}, {Forveille}, \& {Udry}}]{he}
{Gomes da Silva}, J., {Santos}, N.~C., {Bonfils}, X., {et~al.} 2011, \aap, 534, A30, \dodoi{10.1051/0004-6361/201116971}

\bibitem[{{Goodman} \& {Weare}(2010)}]{Goodman2010}
{Goodman}, J., \& {Weare}, J. 2010, Communications in Applied Mathematics and Computational Science, 5, 65, \dodoi{10.2140/camcos.2010.5.65}

\bibitem[{Hobson {et~al.}(2021)Hobson, Brahm, Jordán, Espinoza, Kossakowski, Henning, Rojas, Schlecker, Sarkis, Trifonov, Thorngren, Binnenfeld, Shahaf, Zucker, Ricker, Latham, Seager, Winn, Jenkins, Addison, Bouchy, Bowler, Briegal, Bryant, Collins, Daylan, Grieves, Horner, Huang, Kane, Kielkopf, McLean, Mengel, Nielsen, Okumura, Jones, Plavchan, Shporer, Smith, Tilbrook, Tinney, Twicken, Udry, Unger, West, Wittenmyer, Wohler, Torres, \& Wright}]{wine4}
Hobson, M.~J., Brahm, R., Jordán, A., {et~al.} 2021, The Astronomical Journal, 161, 235, \dodoi{10.3847/1538-3881/abeaa1}

\bibitem[{Hobson {et~al.}(2023)Hobson, Trifonov, Henning, Jordán, Rojas, Espinoza, Brahm, Eberhardt, Jones, Mekarnia, Kossakowski, Schlecker, Pinto, Miranda, Abe, Barkaoui, Bendjoya, Bouchy, Buttu, Carleo, Collins, Colón, Crouzet, Dragomir, Dransfield, Gasparetto, Goeke, Guillot, Günther, Howard, Jenkins, Korth, Latham, Lendl, Lissauer, Mann, Mireles, Ricker, Saesen, Schwarz, Seager, Sefako, Shporer, Stockdale, Suarez, Tan, Triaud, Ulmer-Moll, Vanderspek, Winn, Wohler, \& Zhou}]{TOI-199}
Hobson, M.~J., Trifonov, T., Henning, T., {et~al.} 2023, The Astronomical Journal, 166, 201, \dodoi{10.3847/1538-3881/acfc1d}

\bibitem[{{Holczer} {et~al.}(2016){Holczer}, {Mazeh}, {Nachmani}, {Jontof-Hutter}, {Ford}, {Fabrycky}, {Ragozzine}, {Kane}, \& {Steffen}}]{Holczer2016}
{Holczer}, T., {Mazeh}, T., {Nachmani}, G., {et~al.} 2016, \apjs, 225, 9, \dodoi{10.3847/0067-0049/225/1/9}

\bibitem[{{Holmberg} \& {Madhusudhan}(2024)}]{holmberg:2024}
{Holmberg}, M., \& {Madhusudhan}, N. 2024, \aap, 683, L2, \dodoi{10.1051/0004-6361/202348238}

\bibitem[{{Huang} {et~al.}(2020{\natexlab{a}}){Huang}, {Vanderburg}, {P{\'a}l}, {Sha}, {Yu}, {Fong}, {Fausnaugh}, {Shporer}, {Guerrero}, {Vanderspek}, \& {Ricker}}]{QLPa}
{Huang}, C.~X., {Vanderburg}, A., {P{\'a}l}, A., {et~al.} 2020{\natexlab{a}}, Research Notes of the American Astronomical Society, 4, 204, \dodoi{10.3847/2515-5172/abca2e}

\bibitem[{{Huang} {et~al.}(2020{\natexlab{b}}){Huang}, {Vanderburg}, {P{\'a}l}, {Sha}, {Yu}, {Fong}, {Fausnaugh}, {Shporer}, {Guerrero}, {Vanderspek}, \& {Ricker}}]{QLPb}
---. 2020{\natexlab{b}}, Research Notes of the American Astronomical Society, 4, 206, \dodoi{10.3847/2515-5172/abca2d}

\bibitem[{{Hubbard} \& {Marley}(1989)}]{hubb1989}
{Hubbard}, W.~B., \& {Marley}, M.~S. 1989, \icarus, 78, 102, \dodoi{10.1016/0019-1035(89)90072-9}

\bibitem[{{Jenkins}(2002)}]{Jenkins2002}
{Jenkins}, J.~M. 2002, \apj, 575, 493, \dodoi{10.1086/341136}

\bibitem[{{Jenkins} {et~al.}(2020){Jenkins}, {Tenenbaum}, {Seader}, {Burke}, {McCauliff}, {Smith}, {Twicken}, \& {Chandrasekaran}}]{Jenkins2020}
{Jenkins}, J.~M., {Tenenbaum}, P., {Seader}, S., {et~al.} 2020, {Kepler Data Processing Handbook: Transiting Planet Search}, Kepler Science Document KSCI-19081-003

\bibitem[{{Jenkins} {et~al.}(2016){Jenkins}, {Twicken}, {McCauliff}, {Campbell}, {Sanderfer}, {Lung}, {Mansouri-Samani}, {Girouard}, {Tenenbaum}, {Klaus}, {Smith}, {Caldwell}, {Chacon}, {Henze}, {Heiges}, {Latham}, {Morgan}, {Swade}, {Rinehart}, \& {Vanderspek}}]{SPOC}
{Jenkins}, J.~M., {Twicken}, J.~D., {McCauliff}, S., {et~al.} 2016, in Society of Photo-Optical Instrumentation Engineers (SPIE) Conference Series, Vol. 9913, Software and Cyberinfrastructure for Astronomy IV, ed. G.~{Chiozzi} \& J.~C. {Guzman}, 99133E, \dodoi{10.1117/12.2233418}

\bibitem[{{Jones} {et~al.}(2024){Jones}, {Reinarz}, {Brahm}, {Tala Pinto}, {Eberhardt}, {Rojas}, {Triaud}, {Gupta}, {Ziegler}, {Hobson}, {Jord{\'a}n}, {Henning}, {Trifonov}, {Schlecker}, {Espinoza}, {Torres-Miranda}, {Sarkis}, {Ulmer-Moll}, {Lendl}, {Uzundag}, {Moyano}, {Hesse}, {Caldwell}, {Shporer}, {Lund}, {Jenkins}, {Seager}, {Winn}, {Ricker}, {Burke}, {Figueira}, {Psaridi}, {Al Moulla}, {Mounzer}, {Standing}, {Martin}, {Dransfield}, {Baycroft}, {Dragomir}, {Boyle}, {Suc}, {Mann}, {Timmermans}, {Ducrot}, {Hooton}, {Zu{\~n}iga-Fern{\'a}ndez}, {Sebastian}, {Gillon}, {Queloz}, {Carson}, \& {Lissauer}}]{jones2024}
{Jones}, M.~I., {Reinarz}, Y., {Brahm}, R., {et~al.} 2024, \aap, 683, A192, \dodoi{10.1051/0004-6361/202348147}

\bibitem[{Jordán {et~al.}(2020)Jordán, Brahm, Espinoza, Henning, Jones, Kossakowski, Sarkis, Trifonov, Rojas, Torres, Drass, Nandakumar, Barbieri, Davis, Wang, Bayliss, Bouma, Dragomir, Eastman, Daylan, Guerrero, Barclay, Ting, Henze, Ricker, Vanderspek, Latham, Seager, Winn, Jenkins, Wittenmyer, Bowler, Crossfield, Horner, Kane, Kielkopf, Morton, Plavchan, Tinney, Addison, Mengel, Okumura, Shahaf, Mazeh, Rabus, Shporer, Ziegler, Mann, \& Hart}]{wine2}
Jordán, A., Brahm, R., Espinoza, N., {et~al.} 2020, The Astronomical Journal, 159, 145, \dodoi{10.3847/1538-3881/ab6f67}

\bibitem[{{Kaufer} {et~al.}(1999){Kaufer}, {Stahl}, {Tubbesing}, {N{\o}rregaard}, {Avila}, {Francois}, {Pasquini}, \& {Pizzella}}]{feros}
{Kaufer}, A., {Stahl}, O., {Tubbesing}, S., {et~al.} 1999, The Messenger, 95, 8

\bibitem[{{Kempton} {et~al.}(2018){Kempton}, {Bean}, {Louie}, {Deming}, {Koll}, {Mansfield}, {Christiansen}, {L{\'o}pez-Morales}, {Swain}, {Zellem}, {Ballard}, {Barclay}, {Barstow}, {Batalha}, {Beatty}, {Berta-Thompson}, {Birkby}, {Buchhave}, {Charbonneau}, {Cowan}, {Crossfield}, {de Val-Borro}, {Doyon}, {Dragomir}, {Gaidos}, {Heng}, {Hu}, {Kane}, {Kreidberg}, {Mallonn}, {Morley}, {Narita}, {Nascimbeni}, {Pall{\'e}}, {Quintana}, {Rauscher}, {Seager}, {Shkolnik}, {Sing}, {Sozzetti}, {Stassun}, {Valenti}, \& {von Essen}}]{tsm}
{Kempton}, E. M.~R., {Bean}, J.~L., {Louie}, D.~R., {et~al.} 2018, \pasp, 130, 114401, \dodoi{10.1088/1538-3873/aadf6f}

\bibitem[{{Kley} \& {Nelson}(2012)}]{Kley2012}
{Kley}, W., \& {Nelson}, R.~P. 2012, \araa, 50, 211, \dodoi{10.1146/annurev-astro-081811-125523}

\bibitem[{{Korth} {et~al.}(2024){Korth}, {Chaturvedi}, {Parviainen}, {Carleo}, {Endl}, {Guenther}, {Nowak}, {Persson}, {MacQueen}, {Mustill}, {Cabrera}, {Cochran}, {Lillo-Box}, {Hobbs}, {Murgas}, {Greklek-McKeon}, {Kellermann}, {H{\'e}brard}, {Fukui}, {Pall{\'e}}, {Jenkins}, {Twicken}, {Collins}, {Quinn}, {{\v{S}}ubjak}, {Beck}, {Gandolfi}, {Mathur}, {Deeg}, {Latham}, {Albrecht}, {Barrado}, {Boisse}, {Bouy}, {Delfosse}, {Demangeon}, {Garc{\'\i}a}, {Hatzes}, {Heidari}, {Ikuta}, {Kab{\'a}th}, {Knutson}, {Livingston}, {Martioli}, {Morales-Calder{\'o}n}, {Morello}, {Narita}, {Orell-Miquel}, {Osborne}, {Palakkatharappil}, {Pinter}, {Redfield}, {Relles}, {Schwarz}, {Seager}, {Shporer}, {Skarka}, {Srdoc}, {Stangret}, {Thomas}, {Van Eylen}, {Watanabe}, \& {Winn}}]{toi-1408}
{Korth}, J., {Chaturvedi}, P., {Parviainen}, H., {et~al.} 2024, \apjl, 971, L28, \dodoi{10.3847/2041-8213/ad65fd}

\bibitem[{{Kreidberg}(2015)}]{batman}
{Kreidberg}, L. 2015, Publications of the Astronomical Society of the Pacific, 127, 1161, \dodoi{10.1086/683602}

\bibitem[{{Lee} \& {Peale}(2003)}]{Lee2003}
{Lee}, M., \& {Peale}, S. 2003, \apj, 592, 1201, \dodoi{10.1086/375857}

\bibitem[{{Lee}(2004)}]{Lee2004}
{Lee}, M.~H. 2004, \apj, 611, 517, \dodoi{10.1086/422166}

\bibitem[{{Lee} \& {Peale}(2002)}]{Lee2002}
{Lee}, M.~H., \& {Peale}, S.~J. 2002, \apj, 567, 596, \dodoi{10.1086/338504}

\bibitem[{{Li} {et~al.}(2019){Li}, {Tenenbaum}, {Twicken}, {Burke}, {Jenkins}, {Quintana}, {Rowe}, \& {Seader}}]{Li2019}
{Li}, J., {Tenenbaum}, P., {Twicken}, J.~D., {et~al.} 2019, \pasp, 131, 024506, \dodoi{10.1088/1538-3873/aaf44d}

\bibitem[{{Lightkurve Collaboration} {et~al.}(2018){Lightkurve Collaboration}, {Cardoso}, {Hedges}, {Gully-Santiago}, {Saunders}, {Cody}, {Barclay}, {Hall}, {Sagear}, {Turtelboom}, {Zhang}, {Tzanidakis}, {Mighell}, {Coughlin}, {Bell}, {Berta-Thompson}, {Williams}, {Dotson}, \& {Barentsen}}]{lightkurve}
{Lightkurve Collaboration}, {Cardoso}, J. V. d.~M., {Hedges}, C., {et~al.} 2018, {Lightkurve: Kepler and TESS time series analysis in Python}, Astrophysics Source Code Library, record ascl:1812.013

\bibitem[{{Madhusudhan} {et~al.}(2023){Madhusudhan}, {Sarkar}, {Constantinou}, {Holmberg}, {Piette}, \& {Moses}}]{madhusudhan:2023}
{Madhusudhan}, N., {Sarkar}, S., {Constantinou}, S., {et~al.} 2023, \apjl, 956, L13, \dodoi{10.3847/2041-8213/acf577}

\bibitem[{{Millholland} {et~al.}(2018){Millholland}, {Laughlin}, {Teske}, {Butler}, {Burt}, {Holden}, {Vogt}, {Crane}, {Shectman}, \& {Thompson}}]{GJ876_1}
{Millholland}, S., {Laughlin}, G., {Teske}, J., {et~al.} 2018, \aj, 155, 106, \dodoi{10.3847/1538-3881/aaa894}

\bibitem[{{Miralda-Escud{\'e}}(2002)}]{TTV2}
{Miralda-Escud{\'e}}, J. 2002, \apj, 564, 1019, \dodoi{10.1086/324279}

\bibitem[{{M{\"u}ller} {et~al.}(2024){M{\"u}ller}, {Baron}, {Helled}, {Bouchy}, \& {Parc}}]{muller:2024}
{M{\"u}ller}, S., {Baron}, J., {Helled}, R., {Bouchy}, F., \& {Parc}, L. 2024, \aap, 686, A296, \dodoi{10.1051/0004-6361/202348690}

\bibitem[{Nelder \& Mead(1965)}]{NelderMead}
Nelder, J.~A., \& Mead, R. 1965, Computer Journal, 7, 308

\bibitem[{{Nesvorn{\'y}} {et~al.}(2013){Nesvorn{\'y}}, {Kipping}, {Terrell}, {Hartman}, {Bakos}, \& {Buchhave}}]{KOI-142}
{Nesvorn{\'y}}, D., {Kipping}, D., {Terrell}, D., {et~al.} 2013, \apj, 777, 3, \dodoi{10.1088/0004-637X/777/1/3}

\bibitem[{{Nesvorn{\'y}} {et~al.}(2012){Nesvorn{\'y}}, {Kipping}, {Buchhave}, {Bakos}, {Hartman}, \& {Schmitt}}]{KEPLER46}
{Nesvorn{\'y}}, D., {Kipping}, D.~M., {Buchhave}, L.~A., {et~al.} 2012, Science, 336, 1133, \dodoi{10.1126/science.1221141}

\bibitem[{{Noyes} {et~al.}(1984){Noyes}, {Hartmann}, {Baliunas}, {Duncan}, \& {Vaughan}}]{log}
{Noyes}, R.~W., {Hartmann}, L.~W., {Baliunas}, S.~L., {Duncan}, D.~K., \& {Vaughan}, A.~H. 1984, \apj, 279, 763, \dodoi{10.1086/161945}

\bibitem[{{Paegert} {et~al.}(2021){Paegert}, {Stassun}, {Collins}, {Pepper}, {Torres}, {Jenkins}, {Twicken}, \& {Latham}}]{tic_2}
{Paegert}, M., {Stassun}, K.~G., {Collins}, K.~A., {et~al.} 2021, arXiv e-prints, arXiv:2108.04778, \dodoi{10.48550/arXiv.2108.04778}

\bibitem[{{Panichi} {et~al.}(2018){Panichi}, {Go{\'z}dziewski}, {Migaszewski}, \& {Szuszkiewicz}}]{Kepler30b}
{Panichi}, F., {Go{\'z}dziewski}, K., {Migaszewski}, C., \& {Szuszkiewicz}, E. 2018, \mnras, 478, 2480, \dodoi{10.1093/mnras/sty1071}

\bibitem[{{Paxton} {et~al.}(2011){Paxton}, {Bildsten}, {Dotter}, {Herwig}, {Lesaffre}, \& {Timmes}}]{mesa1}
{Paxton}, B., {Bildsten}, L., {Dotter}, A., {et~al.} 2011, \apjs, 192, 3, \dodoi{10.1088/0067-0049/192/1/3}

\bibitem[{{Paxton} {et~al.}(2013){Paxton}, {Cantiello}, {Arras}, {Bildsten}, {Brown}, {Dotter}, {Mankovich}, {Montgomery}, {Stello}, {Timmes}, \& {Townsend}}]{mesa2}
{Paxton}, B., {Cantiello}, M., {Arras}, P., {et~al.} 2013, \apjs, 208, 4, \dodoi{10.1088/0067-0049/208/1/4}

\bibitem[{Pecaut \& Mamajek(2013)}]{sp_type}
Pecaut, M.~J., \& Mamajek, E.~E. 2013, The Astrophysical Journal Supplement Series, 208, 9, \dodoi{10.1088/0067-0049/208/1/9}

\bibitem[{{Pepe} {et~al.}(2021){Pepe}, {Cristiani}, {Rebolo}, {Santos}, {Dekker}, {Cabral}, {Di Marcantonio}, {Figueira}, {Lo Curto}, {Lovis}, {Mayor}, {M{\'e}gevand}, {Molaro}, {Riva}, {Zapatero Osorio}, {Amate}, {Manescau}, {Pasquini}, {Zerbi}, {Adibekyan}, {Abreu}, {Affolter}, {Alibert}, {Aliverti}, {Allart}, {Allende Prieto}, {{\'A}lvarez}, {Alves}, {Avila}, {Baldini}, {Bandy}, {Barros}, {Benz}, {Bianco}, {Borsa}, {Bourrier}, {Bouchy}, {Broeg}, {Calderone}, {Cirami}, {Coelho}, {Conconi}, {Coretti}, {Cumani}, {Cupani}, {D'Odorico}, {Damasso}, {Deiries}, {Delabre}, {Demangeon}, {Dumusque}, {Ehrenreich}, {Faria}, {Fragoso}, {Genolet}, {Genoni}, {G{\'e}nova Santos}, {Gonz{\'a}lez Hern{\'a}ndez}, {Hughes}, {Iwert}, {Kerber}, {Knudstrup}, {Landoni}, {Lavie}, {Lillo-Box}, {Lizon}, {Maire}, {Martins}, {Mehner}, {Micela}, {Modigliani}, {Monteiro}, {Monteiro}, {Moschetti}, {Murphy}, {Nunes}, {Oggioni}, {Oliveira}, {Oshagh}, {Pall{\'e}}, {Pariani}, {Poretti}, {Rasilla}, {Rebord{\~a}o}, {Redaelli}, {Santana Tschudi},
  {Santin}, {Santos}, {S{\'e}gransan}, {Schmidt}, {Segovia}, {Sosnowska}, {Sozzetti}, {Sousa}, {Span{\`o}}, {Su{\'a}rez Mascare{\~n}o}, {Tabernero}, {Tenegi}, {Udry}, \& {Zanutta}}]{Pepe2021}
{Pepe}, F., {Cristiani}, S., {Rebolo}, R., {et~al.} 2021, \aap, 645, A96, \dodoi{10.1051/0004-6361/202038306}

\bibitem[{{Queloz} {et~al.}(2001){Queloz}, {Henry}, {Sivan}, {Baliunas}, {Beuzit}, {Donahue}, {Mayor}, {Naef}, {Perrier}, \& {Udry}}]{bis}
{Queloz}, D., {Henry}, G.~W., {Sivan}, J.~P., {et~al.} 2001, \aap, 379, 279, \dodoi{10.1051/0004-6361:20011308}

\bibitem[{{Ricker} {et~al.}(2015){Ricker}, {Winn}, {Vanderspek}, {Latham}, {Bakos}, {Bean}, {Berta-Thompson}, {Brown}, {Buchhave}, {Butler}, {Butler}, {Chaplin}, {Charbonneau}, {Christensen-Dalsgaard}, {Clampin}, {Deming}, {Doty}, {De Lee}, {Dressing}, {Dunham}, {Endl}, {Fressin}, {Ge}, {Henning}, {Holman}, {Howard}, {Ida}, {Jenkins}, {Jernigan}, {Johnson}, {Kaltenegger}, {Kawai}, {Kjeldsen}, {Laughlin}, {Levine}, {Lin}, {Lissauer}, {MacQueen}, {Marcy}, {McCullough}, {Morton}, {Narita}, {Paegert}, {Palle}, {Pepe}, {Pepper}, {Quirrenbach}, {Rinehart}, {Sasselov}, {Sato}, {Seager}, {Sozzetti}, {Stassun}, {Sullivan}, {Szentgyorgyi}, {Torres}, {Udry}, \& {Villasenor}}]{TESS}
{Ricker}, G.~R., {Winn}, J.~N., {Vanderspek}, R., {et~al.} 2015, Journal of Astronomical Telescopes, Instruments, and Systems, 1, 014003, \dodoi{10.1117/1.JATIS.1.1.014003}

\bibitem[{{Rivera} {et~al.}(2005){Rivera}, {Lissauer}, {Butler}, {Marcy}, {Vogt}, {Fischer}, {Brown}, {Laughlin}, \& {Henry}}]{GJ876_2}
{Rivera}, E.~J., {Lissauer}, J.~J., {Butler}, R.~P., {et~al.} 2005, \apj, 634, 625, \dodoi{10.1086/491669}

\bibitem[{Schlecker {et~al.}(2020)Schlecker, Kossakowski, Brahm, Espinoza, Henning, Carone, Molaverdikhani, Trifonov, Mollière, Hobson, Jordán, Rojas, Klahr, Sarkis, Bakos, Bhatti, Osip, Suc, Ricker, Vanderspek, Latham, Seager, Winn, Jenkins, Vezie, Villaseñor, Rose, Rodriguez, Rodriguez, Quinn, \& Shporer}]{wine3}
Schlecker, M., Kossakowski, D., Brahm, R., {et~al.} 2020, The Astronomical Journal, 160, 275, \dodoi{10.3847/1538-3881/abbe03}

\bibitem[{{Skilling}(2004)}]{Skilling2004}
{Skilling}, J. 2004, in American Institute of Physics Conference Series, Vol. 735, American Institute of Physics Conference Series, ed. R.~{Fischer}, R.~{Preuss}, \& U.~V. {Toussaint}, 395--405, \dodoi{10.1063/1.1835238}

\bibitem[{{Smith} {et~al.}(2012){Smith}, {Stumpe}, {Van Cleve}, {Jenkins}, {Barclay}, {Fanelli}, {Girouard}, {Kolodziejczak}, {McCauliff}, {Morris}, \& {Twicken}}]{pdcsap3}
{Smith}, J.~C., {Stumpe}, M.~C., {Van Cleve}, J.~E., {et~al.} 2012, \pasp, 124, 1000, \dodoi{10.1086/667697}

\bibitem[{{Speagle}(2020{\natexlab{a}})}]{Speagle2020}
{Speagle}, J.~S. 2020{\natexlab{a}}, \mnras, 493, 3132, \dodoi{10.1093/mnras/staa278}

\bibitem[{{Speagle}(2020{\natexlab{b}})}]{dynesty}
---. 2020{\natexlab{b}}, \mnras, 493, 3132, \dodoi{10.1093/mnras/staa278}

\bibitem[{{Stassun} {et~al.}(2019){Stassun}, {Oelkers}, {Paegert}, {Torres}, {Pepper}, {De Lee}, {Collins}, {Latham}, {Muirhead}, {Chittidi}, {Rojas-Ayala}, {Fleming}, {Rose}, {Tenenbaum}, {Ting}, {Kane}, {Barclay}, {Bean}, {Brassuer}, {Charbonneau}, {Ge}, {Lissauer}, {Mann}, {McLean}, {Mullally}, {Narita}, {Plavchan}, {Ricker}, {Sasselov}, {Seager}, {Sharma}, {Shiao}, {Sozzetti}, {Stello}, {Vanderspek}, {Wallace}, \& {Winn}}]{tic_1}
{Stassun}, K.~G., {Oelkers}, R.~J., {Paegert}, M., {et~al.} 2019, \aj, 158, 138, \dodoi{10.3847/1538-3881/ab3467}

\bibitem[{{Stumpe} {et~al.}(2014){Stumpe}, {Smith}, {Catanzarite}, {Van Cleve}, {Jenkins}, {Twicken}, \& {Girouard}}]{pdcsap2}
{Stumpe}, M.~C., {Smith}, J.~C., {Catanzarite}, J.~H., {et~al.} 2014, \pasp, 126, 100, \dodoi{10.1086/674989}

\bibitem[{{Stumpe} {et~al.}(2012){Stumpe}, {Smith}, {Van Cleve}, {Twicken}, {Barclay}, {Fanelli}, {Girouard}, {Jenkins}, {Kolodziejczak}, {McCauliff}, \& {Morris}}]{pdcsap1}
{Stumpe}, M.~C., {Smith}, J.~C., {Van Cleve}, J.~E., {et~al.} 2012, \pasp, 124, 985, \dodoi{10.1086/667698}

\bibitem[{{Tan} {et~al.}(2013){Tan}, {Payne}, {Lee}, {Ford}, {Howard}, {Johnson}, {Marcy}, \& {Wright}}]{Tan2013}
{Tan}, X., {Payne}, M.~J., {Lee}, M.~H., {et~al.} 2013, \apj, 777, 101, \dodoi{10.1088/0004-637X/777/2/101}

\bibitem[{Tayar {et~al.}(2022)Tayar, Claytor, Huber, \& van Saders}]{Tayar_2022}
Tayar, J., Claytor, Z.~R., Huber, D., \& van Saders, J. 2022, The Astrophysical Journal, 927, 31, \dodoi{10.3847/1538-4357/ac4bbc}

\bibitem[{{Tingley} {et~al.}(2011){Tingley}, {Palle}, {Parviainen}, {Deeg}, {Zapatero Osorio}, {Cabrera-Lavers}, {Belmonte}, {Rodriguez}, {Murgas}, \& {Ribas}}]{Kepler30a}
{Tingley}, B., {Palle}, E., {Parviainen}, H., {et~al.} 2011, \aap, 536, L9, \dodoi{10.1051/0004-6361/201118264}

\bibitem[{Tokovinin(2018)}]{SOAR}
Tokovinin, A. 2018, Publications of the Astronomical Society of the Pacific, 130, 035002, \dodoi{10.1088/1538-3873/aaa7d9}

\bibitem[{{Trifonov}(2019)}]{exostriker}
{Trifonov}, T. 2019, {The Exo-Striker: Transit and radial velocity interactive fitting tool for orbital analysis and N-body simulations}, Astrophysics Source Code Library, record ascl:1906.004

\bibitem[{Trifonov {et~al.}(2021)Trifonov, Brahm, Espinoza, Henning, Jordán, Nesvorny, Dawson, Lissauer, Lee, Kossakowski, Rojas, Hobson, Sarkis, Schlecker, Bitsch, Bakos, Barbieri, Bhatti, Butler, Crane, Nandakumar, Díaz, Shectman, Teske, Torres, Suc, Vines, Wang, Ricker, Shporer, Vanderburg, Dragomir, Vanderspek, Burke, Daylan, Shiao, Jenkins, Wohler, Seager, \& Winn}]{wine5}
Trifonov, T., Brahm, R., Espinoza, N., {et~al.} 2021, The Astronomical Journal, 162, 283, \dodoi{10.3847/1538-3881/ac1bbe}

\bibitem[{{Trifonov} {et~al.}(2021){Trifonov}, {Brahm}, {Espinoza}, {Henning}, {Jord{\'a}n}, {Nesvorny}, {Dawson}, {Lissauer}, {Lee}, {Kossakowski}, {Rojas}, {Hobson}, {Sarkis}, {Schlecker}, {Bitsch}, {Bakos}, {Barbieri}, {Bhatti}, {Butler}, {Crane}, {Nandakumar}, {D{\'\i}az}, {Shectman}, {Teske}, {Torres}, {Suc}, {Vines}, {Wang}, {Ricker}, {Shporer}, {Vanderburg}, {Dragomir}, {Vanderspek}, {Burke}, {Daylan}, {Shiao}, {Jenkins}, {Wohler}, {Seager}, \& {Winn}}]{TOI-2202}
{Trifonov}, T., {Brahm}, R., {Espinoza}, N., {et~al.} 2021, \aj, 162, 283, \dodoi{10.3847/1538-3881/ac1bbe}

\bibitem[{{Trifonov} {et~al.}(2023){Trifonov}, {Brahm}, {Jord{\'a}n}, {Hartogh}, {Henning}, {Hobson}, {Schlecker}, {Howard}, {Reichardt}, {Espinoza}, {Lee}, {Nesvorny}, {Rojas}, {Barkaoui}, {Kossakowski}, {Boyle}, {Dreizler}, {K{\"u}rster}, {Heller}, {Guillot}, {Triaud}, {Abe}, {Agabi}, {Bendjoya}, {Crouzet}, {Dransfield}, {Gasparetto}, {G{\"u}nther}, {Marie-Sainte}, {M{\'e}karnia}, {Suarez}, {Teske}, {Butler}, {Crane}, {Shectman}, {Ricker}, {Shporer}, {Vanderspek}, {Jenkins}, {Wohler}, {Collins}, {Collins}, {Ciardi}, {Barclay}, {Mireles}, {Seager}, \& {Winn}}]{Trifonov2023}
{Trifonov}, T., {Brahm}, R., {Jord{\'a}n}, A., {et~al.} 2023, \aj, 165, 179, \dodoi{10.3847/1538-3881/acba9b}

\bibitem[{Trifonov {et~al.}(2023)Trifonov, Brahm, Jordán, Hartogh, Henning, Hobson, Schlecker, Howard, Reichardt, Espinoza, Lee, Nesvorny, Rojas, Barkaoui, Kossakowski, Boyle, Dreizler, Kürster, Heller, Guillot, Triaud, Abe, Agabi, Bendjoya, Crouzet, Dransfield, Gasparetto, Günther, Marie-Sainte, Mékarnia, Suarez, Teske, Butler, Crane, Shectman, Ricker, Shporer, Vanderspek, Jenkins, Wohler, Collins, Collins, Ciardi, Barclay, Mireles, Seager, \& Winn}]{TOI-2525}
Trifonov, T., Brahm, R., Jordán, A., {et~al.} 2023, The Astronomical Journal, 165, 179, \dodoi{10.3847/1538-3881/acba9b}

\bibitem[{{Twicken} {et~al.}(2018){Twicken}, {Catanzarite}, {Clarke}, {Girouard}, {Jenkins}, {Klaus}, {Li}, {McCauliff}, {Seader}, {Tenenbaum}, {Wohler}, {Bryson}, {Burke}, {Caldwell}, {Haas}, {Henze}, \& {Sanderfer}}]{Twicken2018}
{Twicken}, J.~D., {Catanzarite}, J.~H., {Clarke}, B.~D., {et~al.} 2018, \pasp, 130, 064502, \dodoi{10.1088/1538-3873/aab694}

\bibitem[{{Wisdom} \& {Holman}(1991)}]{Wisdom1991}
{Wisdom}, J., \& {Holman}, M. 1991, \aj, 102, 1528, \dodoi{10.1086/115978}

\bibitem[{{Zechmeister} \& {K{\"u}rster}(2009)}]{Zechmeister2009}
{Zechmeister}, M., \& {K{\"u}rster}, M. 2009, \aap, 496, 577, \dodoi{10.1051/0004-6361:200811296}

\bibitem[{Ziegler {et~al.}(2019)Ziegler, Tokovinin, Briceño, Mang, Law, \& Mann}]{SOAR_TESS}
Ziegler, C., Tokovinin, A., Briceño, C., {et~al.} 2019, The Astronomical Journal, 159, 19, \dodoi{10.3847/1538-3881/ab55e9}

\end{thebibliography}
\bibliographystyle{aasjournal}



\end{document}